\documentstyle{arPrep}
\begin{document}

\input psfig.sty  

\jname{Annu. Rev. Astron. Astrophys.}
\jyear{2002}
\jvol{}

\def\kms {\ km s$^{-1}$}
\def\msol{\ifmmode {\>M_\odot}\else {$M_\odot$}\fi}
\def\cmsq{\ifmmode {\>{\rm\ cm}^2}\else {cm$^2$}\fi}
\def\psqcm{\ifmmode {\>{\rm cm}^{-2}}\else {cm$^{-2}$}\fi}
\def\psqpc{\ifmmode {\>{\rm pc}^{-2}}\else {pc$^{-2}$}\fi}
\def\pcsq{\ifmmode {\>{\rm\ pc}^2}\else {pc$^2$}\fi}
\def\Tkev{\ifmmode{T_{\rm kev}}\else {$T_{\rm keV}$}\fi}
\def\hubunits{\ifmmode {\>{\rm km\ s^{-1}\ Mpc^{-1}}}\else {km
s$^{-1}$ Mpc$^{-1}$}\fi}
\def\gta{\;\lower 0.5ex\hbox{$\buildrel > \over \sim\ $}}
\def\lta{\;\lower 0.5ex\hbox{$\buildrel < \over \sim\ $}}

\def\phiIV{\ifmmode{\varphi_4}\else {$\varphi_4$}\fi}
\def\phiI{\ifmmode{\varphi_i}\else {$\varphi_i$}\fi}

\def\be{\begin{equation}}
\def\ee{\end{equation}}
\def\bea{\begin{eqnarray}}
\def\eea{\end{eqnarray}}
\def\beas{\begin{eqnarray*}}
\def\eeas{\end{eqnarray*}}
\def\gtrapprox{\;\lower 0.5ex\hbox{$\buildrel >\over \sim\ $}}
\def\lessapprox{\;\lower 0.5ex\hbox{$\buildrel < \over \sim\ $}}
\def\deg   {$^\circ$}
\def\Ftwo  {$F_{-21}$}
\def\Pcos  {$\Phi^0$}
\def\Jtwo  {$J_{-21}$}
\def\Fcos  {$F_{-21}^0$}
\def\Jcos  {$J_{-21}^0$}
\def\Em    {${\cal E}_m$}
\def\ALL   {A_{\scriptscriptstyle LL}}
\def\JLL   {J_{\scriptscriptstyle LL}}
\def\nuLL  {\nu_{\scriptscriptstyle LL}}
\def\sigLL {\sigma_{\scriptscriptstyle LL}}
\def\tauLL {\ifmmode{\tau_{\scriptscriptstyle LL}}\else 
           {$\tau_{\scriptscriptstyle LL}$}\fi}
\def\nuOB  {\nu_{\scriptscriptstyle {\rm OB}}}
\def\aB    {\alpha_{\scriptscriptstyle B}}
\def\nH    {n_{\scriptscriptstyle H}}
\def\Em{\ifmmode{{\rm E}_m}\else {{\rm E}$_m$}\fi}
\def\NH{\ifmmode{{\rm N}_{\scriptscriptstyle\rm H}}\else {{\rm N}$_{\scriptscriptstyle\rm H}$}\fi}

\def\Ha    {H$\alpha$}
\def\Hb    {H$\beta$}
\def\HI    {H${\scriptstyle\rm I}$}
\def\HII    {H${\scriptstyle\rm II}$}
\def\eg    {{\it e.g.}}
\def\ie    {{\it i.e.}}
\def\cf    {{\it cf. }}
\def\qv    {{\it q.v. }}
\def\etal  {\ et al.}
\def\kms{\ifmmode {\>{\rm\ km\ s}^{-1}}\else {\ km s$^{-1}$}\fi}

\def\Em{\ifmmode{{\cal E}_m}\else {{\cal E}$_m$}\fi}
\def\Dm{\ifmmode{{\cal D}_m}\else {{\cal D}$_m$}\fi}
\def\fesc{\ifmmode{\hat{f}_{\rm esc}}\else {$\hat{f}_{\rm esc}$}\fi}
\def\fescs{\ifmmode{f_{\rm esc}}\else {$f_{\rm esc}$}\fi}
\def\rsolar{\ifmmode{r_\odot}\else {$r_\odot$}\fi}
\def\emunit{\ifmmode{{\rm cm}^{-6}{\rm\ pc}}\else {
cm$^{-6}$ pc}\fi}
\def\intensity{\ifmmode{{\rm erg\ cm}^{-2}{\rm\ s}^{-1}
      {\rm\ Hz}^{-1}{\rm\ sr}^{-1}}
      \else {erg cm$^{-2}$ s$^{-1}$ Hz$^{-1}$ sr$^{-1}$}\fi}
\def\flux{\ifmmode{{\rm erg\ cm}^{-2}{\rm\ s}^{-1}}\else {erg
cm$^{-2}$ s$^{-1}$}\fi}
\def\fluxdensity{\ifmmode{{\rm erg\ cm^{-2}\ s^{-1}\ Hz^{-1}}}\else {erg
cm$^{-2}$ s$^{-1}$ Hz$^{-1}$}\fi}
\def\phoflux{\ifmmode{{\rm phot\ cm}^{-2}{\rm\ s}^{-1}}\else {phot
cm$^{-2}$ s$^{-1}$}\fi}
\def\phorate{\ifmmode{{\rm phot\ s}^{-1}}\else {phot s$^{-1}$}\fi}

\def\apj{{\it Ap.J.~}}
\def\apjs{{\it Ap.J.Suppl.~}}
\def\apss{{\it Astrophys.Sp.Science~}}
\def\aj{{\it Astron.J.~}}
\def\aph{{\it astro-ph}}
\def\mn{{\it MNRAS~}}
\def\araa{{\it Annu.Rev.Astron.Astrophys.~}}
\def\pasp{{\it PASP.~}}
\def\aaa{{\it Astron.Astrophys.~}}
\def\aaas{{\it Astron.Astrophys.Suppl.~}}
\def\astroph{{\it astro-ph}}
\def\rmp{{\it Rev.Mod.Phys.~}}

\title{\bf The New Galaxy: Signatures of its Formation}

\author{Ken Freeman \affiliation{Mount Stromlo Observatory, Australia National University, Weston Creek, ACT 2611, Australia; kcf@mso.anu.edu.au}
Joss Bland-Hawthorn \affiliation{Anglo-Australian Observatory, 167 Vimiera Road, Eastwood, NSW 2122, Australia; jbh@aao.gov.au}}

\markboth{Freeman \& Hawthorn}{The New Galaxy}

\begin{abstract}

The formation and evolution of galaxies is one of the
great outstanding problems of astrophysics. Within the broad context
of hierachical structure formation, we have only a crude picture
of how galaxies like our own came into existence.  A detailed physical
picture where individual stellar populations can be associated with
(tagged to) elements of the protocloud is far beyond our current
understanding.  Important clues have begun to emerge from both the
Galaxy (near-field cosmology) and the high redshift universe
(far-field cosmology).  Here we focus on the fossil evidence
provided by the Galaxy.  Detailed studies of the Galaxy lie at the
core of understanding the complex processes involved in baryon
dissipation. This is a necessary first step towards achieving a
successful theory of galaxy formation.

\vskip 0.5truein
{\it Key Words: Cosmology, Local Group, Stellar Populations, Stellar Kinematics}
\vspace{.4in}
\end{abstract}

\maketitle

\vspace{.5in}

\section{PROLOGUE}

\subsection{The New Galaxy}

Weinberg (1977) observed that ``the theory of the formation of galaxies
is one of the great outstanding problems of astrophysics, a problem
that today seems far from solution.'' Although the past two decades
have seen considerable progress, Weinberg's assessment remains largely
true.

Eggen, Lynden-Bell and Sandage (1962; ELS) were the first to show that
it is possible to study galactic archaeology using stellar abundances
and stellar dynamics; this is probably the most influential paper on
the subject of galaxy formation.  ELS studied the motions of high
velocity stars and discovered that, as the metal abundance decreases,
the orbit energies and eccentricities of the stars increased while
their orbital angular momenta decreased.  They inferred that the
metal-poor stars reside in a halo that was created during the rapid
collapse of a relatively uniform, isolated protogalactic cloud shortly
after it decoupled from the universal expansion.  ELS is widely viewed
as advocating a smooth monolithic collapse of the protocloud with a
timescale of order $10^8$ years. But Sandage (1990) stresses that this
is an over-interpretation; a smooth collapse was not one of the
inferences they drew from the stellar kinematics.

In 1977, the ELS picture was challenged by Searle (see also Searle \&
Zinn 1978) who noted that Galactic globular clusters have a wide range
of metal abundances essentially independent of radius from the Galactic
Center. They suggested that this could be explained by a halo built up
over an extended period from {\it independent} fragments with masses of
$\sim 10^8$ M$_\odot$.  In contrast, in the ELS picture, the halo
formed in a rapid free-fall collapse. But halo field stars, as well as
globular clusters, are now believed to show an age spread of $2-3$ Ga
(Marquez \& Schuster 1994);  for an alternative view, see Sandage \&
Cacciari (1990). The current paradigm, that the observations argue
for a halo that has built up over a long period from infalling debris,
has developed after many years of intense debate.

This debate parallelled the changes that were taking place in
theoretical studies of cosmology (\eg\ Peebles 1971; Press \& Schecter
1974).  The ideas of galaxy formation via hierarchical aggregation of
smaller elements from the early universe fit in readily with the Searle
\& Zinn view of the formation of the galactic halo from small
fragments. The possibility of identifying debris from these small
fragments was already around in Eggen's early studies of moving groups,
and this is now an active field of research in theoretical and
observational stellar dynamics. It offers the possibility to
reconstruct at least some properties of the protogalaxy and so to
improve our basic understanding of the galaxy formation process.

We can extend this approach to other components of the Galaxy.  We
will argue the importance of understanding the formation of the
galactic disk, because this is where most of the baryons reside.
Although much of the information about the properties of the
protogalactic baryons has been lost in the dissipation that led to the
galactic disk, a similar dynamical probing of the early properties of
the disk can illuminate the formation of the disk, at least back to
the epoch of last significant dissipation. It is also clear that we do
not need to restrict this probing to stellar dynamical techniques. A
vast amount of fossil information is locked up in the detailed stellar
distribution of chemical elements in the various components of the
Galaxy, and we will discuss the opportunities that this offers.

We are coming into a new era of galactic investigation, in which one
can study the fossil remnants of the early days of the Galaxy in a
broader and more focussed way, not only in the halo but throughout the
major luminous components of the Galaxy. This is what we mean by {\it
The New Galaxy}.  The goal of these studies is to reconstruct as much
as possible of the early galactic history. We will review what has
been achieved so far, and point to some of the ways forward.

\subsection{Near-field and far-field cosmology}

What do we mean by the reconstruction of early galactic history? We
seek a detailed physical understanding of the sequence of events which
led to the Milky Way. Ideally, we would want to tag (i.e. associate)
components of the Galaxy to elements of the protocloud -- the baryon
reservoir which fueled the stars in the Galaxy.

From theory, our prevailing view of structure formation relies on a
hierarchical process driven by the gravitational forces of the
large-scale distribution of cold, dark matter (CDM). The CDM paradigm
provides simple models of galaxy formation within a cosmological
context (Peebles 1974; White \& Rees 1978; Blumenthal\etal\ 1984).
N-body and semi-analytic simulations of the growth of structures in the
early universe have been successful at reproducing some of the
properties of galaxies. Current models include gas pressure, metal
production, radiative cooling and heating, and prescriptions for star
formation.

The number density, properties and spatial distribution of dark matter
halos are well understood within CDM (Sheth \& Tormen 1999;
Jenkins\etal\ 2001). However, computer codes are far from producing
realistic simulations of how baryons produce observable galaxies within
a complex hierarchy of dark matter. This a necessary first step towards
a viable theory or a working model of galaxy formation.

In this review, our approach is anchored to observations of the Galaxy,
interpreted within the broad scope of the CDM hierarchy. Many of the
observables in the Galaxy relate to events which occurred long ago, at
high redshift. Fig.~\ref{fig1} shows the relationship between look-back
time and redshift in the context of the $\Lambda$CDM model: the
redshift range (z $\lta$ 6) of discrete sources in contemporary
observational cosmology matches closely the known ages of the oldest
components in the Galaxy. The Galaxy (near-field cosmology) provides a
link to the distant universe (far-field cosmology).

Before we embark on a detailed overview of the relevant data, we give a
descriptive working picture of the sequence of events involved in
galaxy formation. For continuity, the relevant references are given in
the main body of the review where these issues are discussed in more
detail.

\subsection{A working model of galaxy formation}

Shortly after the Big Bang, cold dark matter began to drive baryons
towards local density enhancements. The first stars formed after the
collapse of the first primordial molecular clouds; these stars produced
the epoch of reionization. The earliest recognizable protocloud may
have begun to assemble at about this time.

Within the context of CDM, the dark halo of the Galaxy assembled first,
although it is likely that its growth continues to the present time.
In some galaxies, the first episodes of gas accretion established the
stellar bulge, the central black hole, the first halo stars and the
globular clusters. In the Galaxy and similar systems, the small stellar
bulge may have formed later from stars in the inner disk.

The early stages of the Galaxy's evolution were marked by violent gas
dynamics and accretion events, leading to the high internal densities
of the first globular clusters, and perhaps to the well-known `black
hole mass -- stellar bulge dispersion' relation. The stellar bulge and
massive black hole may have grown up together during this active time.
We associate this era with the `Golden Age', the phase before $z \sim
1$ when star formation activity and accretion disk activity were at
their peak.

At that time, there was a strong metal gradient from the bulge to the
outer halo. The metal enrichment was rapid in the core of the Galaxy
such that, by $z \sim 1$, the mean metallicities were as high as [Fe/H]
$\sim$ -1 or even higher. In these terms, we can understand why the
inner stellar bulge that we observe today is both old and moderately
metal rich. The first halo stars ([Fe/H] $\approx$ -5 to -2.5) formed
over a more extended volume and presumably date back to the earliest
phase of the protocloud. The first globular clusters formed over a
similar volume from violent gas interactions ([Fe/H] $\approx$ -2.5 to
-1.5). We believe now that many of the halo stars and globulars are
remnants of early satellite galaxies which experienced {\it independent}
chemical evolution before being accreted by the Galaxy.

The spread in [Fe/H], and the relative distribution of the chemical
elements, is a major diagnostic of the evolution of each galactic
component. If the initial mass function is constant, the mean
abundances of the different components give a rough indication of the
number of SN~II enrichments which preceded their formation, although we
note that as time passes, an increasing fraction of Fe is produced by
SN~Ia events. For a given parcel of gas in a closed system, only a 
few SN~II events are required to reach [Fe/H] $\approx$ -3, 30 to 100
events to get to [Fe/H] $\approx$ -1.5, and maybe a thousand events to
reach solar metallicities.  We wish to stress that [Fe/H] is not
a clock:  rather it is a measure of supernova occurrences and the depth
of the different potential wells that a given parcel of gas has
explored.

During the latter stages of the Golden Age, most of the baryons began
to settle to a disk for the first time. Two key observations emphasize
what we consider to be the mystery of the main epoch of baryon
dissipation. First, there are no stars with [Fe/H] $<$ -2.2 which
rotate with the disk.  Secondly, despite all the activity associated
with the Golden Age, at least 80\% of the baryons appear to have
settled gradually to the disk over many Ga; this fraction could be as
high as 95\% if the bulge formed after the disk.

About 10\% of the baryons reside in a `thick disk' which has [Fe/H]
$\approx$ -2.2 to -0.5, compared to the younger thin disk with [Fe/H]
$\approx$ -0.5 to +0.3. It is striking how the globular clusters and
the thick disk have similar abundance ranges, although the detailed
abundance distributions are different. There is also a similarity in
age:  globular clusters show an age range of 12 to 14 Ga, and the thick
disk appears to be at least 12 Ga old. Both the thick disk and
globulars apparently date back to the epoch of baryon dissipation
during z $\sim$ 1$-$5.

Fig.~\ref{fig2} summarises our present understanding of the complex 
age$-$metallicity distribution for the various components of the 
Galaxy.

It is a mystery that the thick disk and the globulars should have
formed so early {\it and} over such a large volume from material which was
already enriched to [Fe/H] $\sim$ -2. Could powerful winds from the
central starburst in the evolving core have distributed metals
throughout the inner protocloud at about that time?

Finally, we emphasize again that 90\% of the disk baryons have settled 
quiescently to the thin disk since z $\sim$ 1.

\subsection{Timescales and fossils}

The oldest stars in our Galaxy are of an age similar to the look-back
time of the most distant galaxies in the Hubble Deep Field
(Fig.~\ref{fig1}).  For the galaxies, the cosmological redshift
measured from galaxy spectra presently takes us to within 5\% of the
origin of cosmic time.  For the stars, their upper atmospheres provide
fossil evidence of the available metals at the time of formation.  The
old Galactic stars and the distant galaxies provide a record of
conditions at early times in cosmic history, and both harbor clues to
the sequence of events which led to the formation of galaxies like the
Milky Way.

The key timescale provided by far-field cosmology is the look-back time
with the prospect of seeing galaxies at an earlier stage in their evolution.
However, this does {\it not} imply that these high-redshift objects are
unevolved. We know that the stellar cores of galaxies at the highest
redshifts ($z\sim 5$) observed to date exhibit solar metallicities, and
therefore appear to have undergone many cycles of star formation
(Hamann \& Ferland 1999). Much of the light we detect from the early
universe probably arises from the chemically and dynamically evolved
cores of galaxies.

Near-field cosmology provides a dynamical timescale, $\tau_D \sim
(G\rho)^{-\frac{1}{2}}$, where $\rho$ is the mean density of the
medium.  The dynamical timescale at a radial distance of 100 kpc is of
order several Ga, so the mixing times are very long. Therefore,
on larger scales, we can expect to find dynamical and chemical traces
of past events, even where small dynamical systems have long since merged
with the Galaxy.

We note that the CDM hierarchy reflects a wide range of dynamical
timescales, such that different parts of the hierarchy may reveal
galaxies in different stages of evolution.  In this sense, the
hierarchy relates the large-scale density to the morphology and
evolution of its individual galaxies; this is the so-called
`morphology-density relation' (Dressler 1980; Hermit\etal\ 1996;
Norberg\etal\ 2001). Over a large enough ensemble of galaxies, taken
from different regions of the hierarchy, we expect different
light-weighted age distributions because one part of the hierarchy is
more evolved than another.  In other words, the evolution of
small-scale structure (individual galaxies) must at some level relate
to the environment on scales of 10 Mpc or more.

The near field also provides important evolutionary timescales for
individual stars and groups of stars (see ``Stellar age dating''
below). Individual stars can be dated with astero-seismology
(Christensen-Dalsgaard 1986; Gough 2001) and nucleo-cosmochronology
(Fowler \& Hoyle 1960; Cowan\etal\ 1997). Strictly speaking,
nucleo-cosmochronology dates the elements rather than the stars. Coeval
groups of stars can be aged from the main-sequence turn-off or from the
He-burning stars in older populations (Chaboyer 1998).  Furthermore,
the faint end cut-off of the white dwarf luminosity function provides
an important age constraint for older populations (Oswalt\etal\ 1996).
Presently, the aging methods are model dependent.

\subsection{Goals of near-field cosmology}

We believe that the major goal of near-field cosmology is to tag
individual stars with elements of the protocloud. Some integrals of
motion are likely to be preserved while others are scrambled by
dissipation and violent relaxation. We suspect that complete tagging is
impossible. However, some stars today may have some integrals
of motion which relate to the protocloud at the epoch of last
dissipation (see ``Zero order signatures -- information preserved since
dark matter virialized'' below).

As we review, different parts of the Galaxy have experienced
dissipation and phase mixing to varying degrees. The disk, in contrast
to the stellar halo, is a highly dissipated structure. The bulge may be
only partly dissipated. To what extent can we unravel the events that
produced the Galaxy as we see it today?  Could some of the residual
inhomogeneities from prehistory have escaped the dissipative process at
an early stage?

Far field cosmology currently takes us back to the `epoch of last
scattering' as seen in the microwave background. Cosmologists would
like to think that some vestige of information has survived from
earlier times (cf. Peebles, Seager \& Hu 2000). In the same spirit, we
can hope that fossils remain from the `epoch of last dissipation',
\ie\ the main epoch of baryon dissipation that occurred as the disk was
being assembled.

To make a comprehensive inventory of surviving inhomogeneities would
require a vast catalog of stellar properties that is presently out of
reach (Bland-Hawthorn 2002).  The Gaia space astrometry mission
(Perryman\etal\ 2001), set to launch at the end of the decade, will
acquire detailed phase space coordinates for about one billion stars,
within a sphere of diameter 20 kpc (the Gaiasphere).  In ``The
Gaiasphere and the limits of knowledge'' below, we look forward to a
time when all stars within the Gaiasphere have complete chemical
abundance measurements (including all heavy metals).  Even with such a
vast increase in information, there may exist fundamental $-$ but
unproven $-$ limits to unravelling the observed complexity.

The huge increase in data rates from ground-based and space-based
observatories has led to an explosion of information.  Much of this
information from the near field is often dismissed as `weather' or
unimportant detail.  But in fact fundamental clues are already
beginning to emerge.  In what is now a famous discovery, a large
photometric and kinematic survey of bulge stars revealed the presence
of the disrupting Sgr dwarf galaxy (Ibata \etal\ 1994), now seen over a
large region of sky and in a variety of populations (see ``Structures
in phase space'' below).  Perhaps the most important example arises
from the chemical signatures seen in echelle spectroscopy of bulge,
thick disk and halo stars. In ``Epilogue: challenges for the future'',
we envisage a time when the analysis of thousands of spectral lines for
a vast number of stars will reveal crucial insights into the sequence
of events early in the formation of the Galaxy.

In this review, we discuss fossil signatures in the Galaxy. A key aspect
of fossil studies is a reliable time sequence. In ``Stellar age dating,''
we discuss methods for age-dating individual stars and coeval groups of
stars.  In  ``Structure of the Galaxy,'' we describe the main components
of the Galaxy.  In ``Signatures of galaxy formation,'' we divide the
fossil signatures of galaxy formation into three parts: zero order
signatures that preserve information since dark matter virialized;
first order signatures that preserve information since the main epoch
of baryon dissipation; second order signatures that arise from major
processes involved in subsequent evolution. In ``The Gaiasphere and the
limits of knowledge,'' we look forward to a time when it is possible to
measure ages, phase space coordinates and chemical properties for a vast
number of stars in the Galaxy.  Even then, what are the prospects for
unravelling the sequence of events that gave rise to the Milky Way? We 
conclude with some experimental challenges for the future.

\section{STELLAR AGE DATING}

Nucleo-cosmochronology (or cosmochronometry), or the aging of the
elements through radioactive decay, has a long history (Rutherford
1904; Fowler \& Hoyle 1960; Butcher 1987). A related technique is
widely used in solar system geophysics. Independent schemes have aged
the oldest meteorites at 4.53$\pm$0.04 Ga (Guenther \& Demarque 1997;
Manuel 2000). The small uncertainties reflect that the age dating is
direct. Element pairs like Rb and Sr are chemically distinct and freeze
out during solidification into different crystalline grains. The
isotope $^{87}$Rb decays into $^{87}$Sr which can be compared to
$^{86}$Sr, a non-radiogenic isotope, measured from a control sample of
Sr-rich grains.  This provides a direct measure of the fraction of a
$^{87}$Rb half-life ($\tau_{1/2} = 47.5$ Ga) since the meteorite
solidified.

It appears that, until we have a precise understanding of BBNS and the
early chemical evolution history of the Galaxy, geophysical precision
will not be possible for stellar ages. The major problem is that, as
far as we know, there is no chemical differentiation which requires
that we know precisely how much of each isotope was originally
present.  Modern nucleo-cosmochronology compares radioactive isotope
strengths to a stable r-process element (\eg\ Nd, Eu, La, Pt).  The
thorium method ($^{232}$Th, $\tau_{1/2} = 14.0$ Ga) was first applied
by Butcher (1987) and refined by Pagel (1989).  Other radioactive
chronometers include $^{235}$U ($\tau_{1/2} = 0.70$ Ga) and $^{238}$U
($\tau_{1/2} = 4.47$ Ga) although Yokoi\etal\ (1983) have expressed
concerns about their use (cf. Cayrel\etal\ 2001).  Arnould \& Goriely
(2001) propose that the isotope pair $^{187}$Re$-^{187}$Os ($\tau_{1/2}
= 43.5$ Ga) may be better suited for future work.

With the above caveats, we point out that several groups are now
obtaining exquisite high-resolution data on stars with enhanced
r-process elements (Cayrel\etal\ 2001; Sneden\etal\ 2000; 
Burris\etal\ 2000; Westin\etal\ 2000; Johnson \& Bolte 2001; 
Cohen\etal\ 2002; Hill\etal\ 2002).  For a subset of these
stars, radioactive ages have been derived (Truran\etal\ 2001)
normalized to the heavy element abundances observed in meteorites.

There are few other direct methods for deriving ages of {\it
individual} stars.  A promising field is asteroseismology which relies
on the evolving mean molecular weight in stellar cores
(Christensen-Dalsgaard 1986; Ulrich 1986; Gough 1987; Guenther 1989).
Gough (2001) has determined 4.57$\pm$0.12~Ga for the Sun which should
be compared with the age of meteorites quoted above. The Eddington
satellite under consideration by ESA for launch at the end of the
decade proposes to use stellar oscillations to age 50,000 main sequence
stars (Gimenez \& Favata 2001).

It has long been known that disk stars span a wide range of ages from
the diversity of main-sequence stars.  Edvardsson\etal\ (1993) derived
precise stellar evolution ages for nearby individual post main-sequence
F stars using Str\"omgren photometry, and showed that the stars in the
Galactic disk exhibit a large age spread with ages up to roughly 10~Ga
(Fig.~\ref{fig2}). Using the `inverse age-luminosity relation' for RR
Lyrae stars, Chaboyer\etal\ (1996) found that the oldest globular
clusters are older than 12 Ga with 95\% confidence, with a best
estimate of 14.6$\pm$1.7 Ga (Chaboyer 1998).  But Hipparcos appears to
show that the RR Lyr distances are underestimated leading to a downward
revision of the cluster ages:  $8.5-13.3$ Ga (Gratton\etal\ 1997);
$11-13$ Ga (Reid 1998); $10.2-12.8$ Ga (Chaboyer\etal\ 1998).  For a
coeval population (\eg\ open and globular clusters), isochrone fitting
is widely used. The ages of the galactic halo and globular clusters,
when averaged over eight independent surveys, lead to 12.2$\pm 0.5$ Ga
(Lineweaver 1999).

Other traditional methods rely on aging a population of stars that are
representative of a particular component of the Galaxy. For example,
Gilmore \etal\ (1989) use the envelope of the distribution in a
color-abundance plane to show that all stars more metal poor than
[Fe/H] $=$ -0.8 are as old as the globular clusters.  Similarly, the
faint end of the white dwarf luminosity function is associated with the
coolest, and therefore the oldest, stars (Oswalt\etal\ 1996). The
present estimate for the age of the old thin disk population when
averaged over five independent surveys is $8.7\pm 0.4$ Ga (Lineweaver
1999), although Oswalt\etal\ argue for $9.5^{+1.1}_{-0.8}$ Ga.

For a world model with ($\Omega_\Lambda = 0.7$, $\Omega_m = 0.3$), the
Big Bang occurred 14~Ga ago (Efstathiou\etal\ 2002) -- in our view,
there is no compelling evidence for an age crisis from a comparison of
estimates in the near and far field.  But the inaccuracy of age dating
relative to an absolute scale does cause problems.  At present, the
absolute ages of the oldest stars cannot be tied down to better than
about 2~Ga, the time elapsed between $z=6$ and $z=2$. This is a
particular handicap to identifying specific events in the early
Universe from the stellar record.
\section{STRUCTURE OF THE GALAXY}

Like most spiral galaxies, our Galaxy has several recognizable
structural components that probably appeared at different stages in
the galaxy formation process. These components will retain different
kinds of signatures of their formation. We will describe these
components in the context of other disk galaxies, and use images of
other galaxies in Fig.~\ref{fig3} to illustrate the components.

\subsection{The Bulge}

First compare images of M104 and IC 5249 (Figs.~\ref{fig3}(c,g)): these 
are extreme examples of
galaxies with a large bulge and with no bulge. Large bulges like that
of M104 are structurally and chemically rather similar to elliptical
galaxies: their surface brightness distribution follows an $r^{1/4}$
law (\eg\ Pritchet \& van den Bergh 1994) and they show similar
relations of [Fe/H] and [Mg/Fe] with absolute magnitude (\eg\ Jablonka
\etal\ 1996). These properties lead to the view that the
large bulges formed rapidly. The smaller bulges are often boxy in
shape, with a more exponential surface brightness distribution (\eg\
Courteau \etal\ 1996). The current belief is that they may
have arisen from the stellar disk through bending mode instabilities.

Spiral bulges are usually assumed to be old but this is poorly known,
even for the Galaxy. The presence of bulge RR Lyrae stars indicates
that at least some fraction of the galactic bulge is old (Rich 2001).
Furthermore, the color-magnitude diagrams for galactic bulge stars show
that the bulge is predominantly old.  McWilliam \& Rich (1994) measured
[Fe/H] abundances for red giant stars in the bulge of the Milky Way.
They found that, while there is a wide spread, the abundances ([Fe/H]
$\approx$ -0.25) are closer to the older stars of the metal rich disk
than to the very old metal poor stars in the halo and in globular
clusters, in agreement with the abundances of planetary nebulae in the
Galactic bulge (\eg\ Exter\etal\ 2001).

The COBE image of the Milky Way (Fig.~\ref{fig3}(b)) shows a modest
somewhat boxy bulge, typical of an Sb to Sc spiral. Fig.~\ref{fig3}(d)
shows a more extreme example of a boxy/peanut bulge. If such bulges do
arise via instabilities of the stellar disk, then much of the
information that we seek about the state of the early galaxy would have
been lost in the processes of disk formation and subsequent bulge
formation.  Although most of the more luminous disk galaxies have
bulges, many of the fainter disk galaxies do not. Bulge formation is
not an essential element of the formation processes of disk galaxies.

\subsection{The Disk} Now look at the disks of these galaxies. The
exponential thin disk, with a vertical scale height of about 300 pc,
is the most conspicuous component in edge-on disk galaxies like NGC
4762 and IC 5249 (Figs.~\ref{fig3}(e,g)). The thin disk is believed to 
be the end product of the
quiescent dissipation of most of the baryons and contains almost all
of the baryonic angular momentum.  For the galactic disk, which is
clearly seen in the COBE image in Fig.~\ref{fig3}(b), we know from 
radioactive dating, white dwarf cooling and isochrone estimates for 
individual evolved stars and open clusters that the oldest disk stars 
have ages in the range 10 to 12 Ga (see ``Stellar age dating'' above).

The disk is the defining stellar component of disk galaxies, and
understanding its formation is in our view the most important goal of
galaxy formation theory.  Although much of the information about the
pre-disk state of the baryons has been lost in the dissipative
process, some tracers remain, and we will discuss them in the next
section. 

Many disk galaxies show a second fainter disk component with a larger
scale height (typically about 1 kpc); this is known as the thick disk.
Deep surface photometry of IC 5249 shows only a very faint thick disk
enveloping the bright thin disk (Abe \etal\ 1999): compare
Figs.~\ref{fig3}(g,h). In the edge-on S0 galaxy NGC 4762, we see a much
brighter thick disk around its very bright thin disk (Tsikoudi 1980):
the thick disk is easily seen by comparing Figs.~\ref{fig3}(e,f).  The
Milky Way has a significant thick disk (Gilmore \& Reid 1983):  its
scale height ($\sim 1$ kpc) is about three times larger than the scale
height of the thin disk, its surface brightness is about 10\% of the
thin disk's surface brightness, its stellar population appears to be
older than about 12 Ga, and its stars are significantly more metal poor
than the stars of the thin disk. The galactic thick disk is currently
believed to arise from heating of the early stellar disk by accretion
events or minor mergers (see ``Disk heating by accretion'' below).

The thick disk may be one of the most significant components for
studying signatures of galaxy formation because it presents a `snap
frozen' relic of the state of the (heated) early disk. Although some
apparently pure-disk galaxies like IC 5249 do have faint thick disks,
others do not (Fry\etal\ 1999): these pure-disk galaxies show no
visible components other than the thin disk.  As for the bulge,
formation of a thick disk is not an essential element of the galaxy
formation process. In some galaxies the dissipative formation of the
disk is clearly a very quiescent process.

\subsection{The Stellar Halo} There are two further components that
are not readily seen in other galaxies, and are shown schematically in
Fig.~\ref{fig3}(a). The first is the metal-poor stellar halo,
well-known in the Galaxy as the population containing the metal-poor
globular clusters and field stars. Its mass is only about 1\% of the
total stellar mass (about $10^9 M_\odot$: \eg\ Morrison 1993).  The
surface brightness of the galactic halo, if observed in other galaxies,
would be too low to detect from its diffuse light. It can be seen in
other galaxies of the Local Group in which it is possible to detect the
individual evolved halo stars.  The metal-poor halo of the Galaxy is
very interesting for galaxy formation studies because it is so old:
most of its stars are probably older than 12 Ga and are probably among
the first galactic objects to form.  The galactic halo has a power law
density distribution $\rho \propto r^{-3.5}$ although this appears to
depend on the stellar population (Vivas\etal\ 2001, Chiba \& Beers
2000).  Unlike the disk and bulge, the angular momentum of the halo is
close to zero (\eg\ Freeman 1987), and it is supported almost entirely
by its velocity dispersion; some of its stars are very energetic,
reaching out to at least 100 kpc from the galactic center (\eg\ Carney
\etal\ 1990).

The current view is that the galactic halo formed at least partly
through the accretion of small metal-poor satellite galaxies which
underwent some independent chemical evolution before being accreted by
the Galaxy (Searle \& Zinn 1978; Freeman 1987).  Although we do
still see such accretion events taking place now, in the apparent tidal
disruption of the Sgr dwarf (Ibata \etal\ 1995), most of
them must have occurred long ago.  Accretion of satellites would
dynamically heat the thin disk, so the presence of a dominant thin disk
in the Galaxy means that most of this halo-building accretion probably
predated the epoch of thick disk formation $\sim 12$ Ga ago. We can
expect to see dynamically unmixed residues or fossils of at least some
of these accretion events (\eg\ Helmi \& White 1999).

Of all the galactic components, the stellar halo offers the best
opportunity for probing the details of its formation. There is a real
possibility to identify groups of halo stars that originate from common
progenitor satellites (Eggen 1977; Helmi \& White 1999;
Harding\etal\ 2001; Majewski\etal\ 2000).  However, if the accretion
picture is correct, then the halo is just the stellar debris of small
objects accreted by the Galaxy early in its life.  Although it may be
possible to unravel this debris and associate individual halo stars
with particular progenitors, this may tell us more about the early
chemical evolution of dwarf galaxies than about the basic issues of
galaxy formation. We would argue that the thin disk and thick disk of
our Galaxy retain the most information about how the Galaxy formed. On
the other hand, we note that current hierarchical CDM simulations
predict many more satellites than are currently observed.  It would
therefore be very interesting to determine the number of satellites
that have already been accreted to form the galactic stellar halo.

We should keep in mind that the stellar halo accounts for only a tiny
fraction of the galactic baryons and is dynamically distinct from the
rest of the stellar baryons.  We should also note that the stellar halo
of the Galaxy may not be typical: the halos of disk galaxies are quite
diverse.  The halo of M31, for example, follows the $r^{1/4}$ law
(Pritchet \& van den Bergh 1994) and is much more metal rich in the
mean than the halo of our Galaxy (Durrell \etal\ 2001) although it 
does have stars that are very metal weak. It should
probably be regarded more as the outer parts of a large bulge than as a
distinct halo component.  For some other disk galaxies, like the LMC, a
metal-poor population is clearly present but may lie in the disk rather
than in a spheroidal halo.

\subsection{The Dark Halo}

The second inconspicuous component is the dark halo, which is detected
only by its gravitational field. The dark halo contributes at least
90\% of the total galactic mass and its $\rho \sim r^{-2}$ density
distribution extends to at least 100 kpc (\eg\ Kochanek 1996). It is
believed to be spheroidal rather than disklike (Cr\'ez\'e\etal\ 1998;
Ibata\etal\ 2001$b$; see Pfenniger \etal\ 1994 for a contrary view). In
the current picture of galaxy formation, the dark halo plays a very
significant role. The disk is believed to form dissipatively within the
potential of the virialized spheroidal halo which itself formed through
the fairly rapid aggregation of smaller bodies.

CDM simulations suggest that the halo may still be strongly substructured
(see ``Signatures of the CDM hierarchy'' below). If this is correct,
then the lumpy halo would continue to influence the evolution of the
galactic disk, and the residual substructure of the halo is a fossil
of its formation. If the dark matter is grainy, it may be possible to
study the dynamics of this substructure through pixel lensing of the
light of background galaxies (Widrow \& Dubinski 1998; Lewis\etal\ 2000).
Another possibility is to look for the signatures of substructure around
external galaxies in gravitationally lensed images of background quasars
(Metcalf 2001; Chiba 2001). The dispersal of tidal tails from globular
clusters appears to be sensitive to halo substructure (Ibata\etal\
2001$b$), although this is not the case for dwarf galaxies (Johnston,
Spergel \& Haydn 2002).

\smallskip
Within the limitations mentioned above, each of these distinct
components of the Galaxy preserves signatures of its past and so 
gives insights into the galaxy formation process. We now discuss 
these signatures.

\section{SIGNATURES OF GALAXY FORMATION}

Our framework is that the Galaxy formed through hierarchical aggregation.
We identify three major epochs in which information about the 
proto-hierarchy is lost:

(i) the dark matter virializes $-$ this could be a time of intense
    star formation but need not be, as evidenced by the existence
    of very thin pure disk galaxies;

(ii) the baryons dissipate to form the disk and bulge;

(iii) an ongoing epoch of formation of objects within the disk and 
   accretion of objects from the environment of the galaxy, both 
   leaving some long-lived relic.

\medskip
We classify signatures relative to these three epochs. The role of the
environment is presently difficult to categorize in this way.
Environmental influences must be operating across all of our
signature classes.

\subsection{\bf Zero order signatures -- information preserved since 
dark matter virialized}

\subsubsection{Introduction}

During the virialization phase, a lot of information about the local
hierarchy is lost: this era is dominated by merging and violent
relaxation. The total dark and baryon mass are probably roughly
conserved, as is the angular momentum of the region of the hierarchy
that went into the halo. The typical density of the environment is also
roughly conserved:  although the structure has evolved through
merging and relaxation, a low density environment remains a low density
environment (see White 1996 for an overview).

\subsubsection{Signatures of the environment}

The local density of galaxies (and particularly the number of small
satellite systems present at this epoch) affects the incidence of later
interactions. For the Local Group, the satellite numbers appear to be
lower than expected from CDM (Moore\etal\ 1999; Klypin\etal\ 1999).
However there is plenty of evidence for past and ongoing
accretion of small objects by the Milky Way and M31 (Ibata 
\etal\ 1995; Ibata\etal\ 2001$b$).

The thin disk component of disk galaxies settles dissipatively in the
potential of the virialised dark halo (\eg\ Fall \& Efstathiou 1980).
The present morphology of the thin disk depends on the numbers of small
galaxies available to be accreted: a very thin disk is an indication of
few accretion events (dark or luminous) after the epoch of disk
dissipation and star formation (\eg\ Freeman 1987; Quinn \etal\ 1993; 
Walker \etal\ 1996).  The formation of the
thick disk is believed to be associated with a discrete event that
occurred very soon after the disk began to settle, at a time when about
10\% of the stars of the disk had already formed.  In a low density
environment, without such events, thick disk formation may not occur.
Since the time of thick disk formation, the disk of the Galaxy appears
to have been relatively undisturbed by accretion events.  This is
consistent with the observation that less than 10\% of the metal-poor
halo comes from recent accretion of star forming satellites (Unavane
\etal\ 1996).

The existence and structure of the metal-poor stellar halo of the
Galaxy may depend on accretion of small objects. This accretion
probably took place after the gaseous disk had more or less settled --
the disk acts as a resonator for the orbit decay of the small objects.
So again the environment of our proto-galaxy may have a strong
signature in the very existence of the stellar halo, and certainly in
its observed substructure. We would not expect to find a stellar halo
encompassing pure disk galaxies, consistent with the limited evidence
now available (Freeman \etal\ 1983; Schommer\etal\ 1992).

\subsubsection{Signatures of global quantities}

During the process of galaxy formation, some baryons are lost to ram
pressure stripping and galactic winds. Most of the remaining baryons
become the luminous components  of the galaxy.  The total angular
momentum $J$ of the dark halo may contribute to its shape, which in
turn may affect the structure of the disks. For example, warps may be
associated with misalignment of the angular momentum of the dark and
baryonic components. The dark halo may have a rotating triaxial
figure:  the effect of a rotating triaxial dark halo on the
self-gravitating disk has not yet been seriously investigated (see
Bureau\etal\ 1999).

The binding energy $E$ at the epoch of halo virialization affects the
depth of the potential well and hence the characteristic velocities in
the galaxy. It also affects the parameter 
$\lambda = J|E|^{\frac{1}{2}}G^{-1}M^{-\frac{5}{2}}$, where $M$ is the
total mass: $\lambda$ which critical for determining the gross nature
of the galactic disk as a high or low surface brightness system
(\eg\ Dalcanton \etal\ 1997).

The relation between the specific angular momentum $J/M$ and the total
mass $M$ (Fall 1983) of disk galaxies is well reproduced by simulations
(Zurek \etal\ 1988).  Until recently, ellipticals and disk galaxies
appeared to be segregated in the Fall diagram: from the slow rotation
of their inner regions, estimates of the $J/M$ ratios for ellipticals
were about 1 dex below those for the spirals.   More recent work
(\eg\ Arnaboldi\etal\ 1994) shows that much of the angular momentum in
ellipticals appears to reside in their outer regions, so ellipticals
and spirals do have similar locations in the Fall diagram.  Internal
redistribution of angular momentum has clearly occurred in the ellipticals
(Quinn \& Zurek 1988).

The remarkable Tully-Fisher law (1977) is a correlation between the HI
line-width and the optical luminosity of disk galaxies. It appears to
relate the depth of the potential well and the baryonic mass (McGaugh
\etal\ 2000). Both of these quantities are probably roughly conserved
after the halo virialises, so the Tully-Fisher law should be regarded
as a zero-order signature of galaxy formation. The likely connecting
links between the (dark) potential well and the baryonic mass are (i) a
similar baryon/dark matter ratio from galaxy to galaxy, and (ii) an
observed Faber-Jackson law for the dark halos, of the form $M \propto
\sigma^4$:  \ie\ surface density independent of mass for the dark halos
(J. Kormendy \& K.C. Freeman, in preparation).

\subsubsection{Signatures of the internal distribution of specific 
angular momentum}

The internal distribution of specific angular momentum ${\cal M}(h)$ of
the baryons (\ie\ the mass with specific angular momentum $< h$)
largely determines the shape of the surface brightness distribution of
the disk rotating in the potential well of the dark matter.  Together
with ${\cal M}(h)$, the total angular momentum and mass of the baryons
determine the scale length and scale surface density of the disk.
Therefore the distribution of total angular momentum and mass for
protodisk galaxies determines the observed distribution of the scale
length and scale surface density for disk galaxies (Freeman 1970; de
Jong \& Lacey 2000).

Many studies have assumed that ${\cal M}(h)$ is conserved through the
galaxy formation process, most notably Fall \& Efstathiou (1980). It is
not yet clear if this assumption is correct. Conservation of the
internal distribution of specific angular momentum ${\cal M}(h)$ is a
much stronger requirement than the conservation of the {\it total}
specific angular momentum $J/M$.  Many processes can cause the internal
angular momentum to be redistributed, while leaving the $J/M$ ratio
unchanged. Examples include the effects of bars, spiral structure
(Lynden-Bell \& Kalnajs 1972) and internal viscosity (Lin \& Pringle
1987).

The maximum specific angular momentum $h_{\rm max}$ of the baryons may
be associated with the truncation of the optical disk observed at about
four scale lengths (de Grijs\etal\ 2001; Pohlen\etal\ 2000).  This needs
more investigation.  The truncation of disks could be an important
signature of the angular momentum properties of the early protocloud,
but it may have more to do with the critical density for star formation
or the dynamical evolution of the disk. Similarly, in galaxies with
very extended HI, the edge of the HI distribution may give some measure
of $h_{\rm max}$ in the protocloud. On the other hand, it may be that
the outer HI was accreted subsequent to the formation of the stellar
disk (van der Kruit 2002), or that the HI edge may just represent the
transition to an ionized disk (Maloney 1993).

This last item emphasizes the importance of understanding what is going
on in the outer disk. The outer disk offers some potentially important
diagnostics of the properties of the protogalaxy. At present there are
too many uncertainties about significance of (a) the various cutoffs in
the light and HI distributions, (b) the age gradient seen by Bell \&
de Jong (2000) from integrated light of disks but not by Friel (1995)
for open clusters in the disk of the Galaxy, and (c) the outermost disk
being maybe younger but not `zero age', which means that there is no
real evidence that the disk is continuing to grow radially. It is
possible that the edge of the disk has something to do with angular
momentum of baryons in the protocloud or with disk formation process,
so it may be a useful zero-order or first-order signature.

\subsubsection{Signatures of the CDM hierarchy}

CDM predicts a high level of substructure which is in apparent conflict
with observation.  Within galaxies, the early N-body simulations
appeared to show that substructure with characteristic velocities in
the range $10 < V_c < 30$\kms\ would be destroyed by merging and
virialization of low mass structures (Peebles 1970; White 1976; White
\& Rees 1978). It turned out that the lack of substructure was an
artefact of the inadequate spatial and mass resolution
(Moore\etal\ 1996). Current simulations reveal 500 or more low mass
structures within 300 kpc of an L$_*$ galaxy's sphere of influence
(Moore\etal\ 1999; Klypin\etal\ 1999).  This is an order of magnitude
larger than the number of low mass satellites in the Local Group. Mateo
(1998) catalogues about 40 such objects and suggests that, at most, we
are missing a further $15-20$ satellites at low galactic latitude.
Kauffmann\etal\ (1993) were the first to point out the `satellite
problem' and suggested that the efficiency of dynamical friction might
be higher than usually quoted.  However, without recourse to fine
tuning, this would remove essentially all of the observed satellites in
the Local Group today.

Since the emergence of the CDM paradigm, an inevitable question is
whether a `basic building block' can be recognized in the near field.
Moore\etal\ emphasize the self-similar nature of CDM sub-clustering and
point to the evidence provided by the mass spectrum of objects in rich
clusters, independent of the N-body simulations.  The lure of finding a
primordial building block in the near field has prompted a number of
tests. If the dark mini-halos comprise discrete sources, it should be
possible to detect microlensing towards a background galaxy (see
``The dark halo'' above).

The satellite problem appears to be a fundamental prediction of CDM in
the non-linear regime.  Alternative cosmologies have been suggested
involving the reduction of small-scale power in the initial mass power
spectrum (Kamionkowski \& Liddle 2000), warm dark matter (Hogan \&
Dalcanton 2001; White \& Croft 2000; Colin\etal\ 2000), or strongly
self-interacting dark matter (Spergel \& Steinhardt 2000). Several
authors have pointed out that some of the direct dark-matter detection
experiments are sensitive to the details of the dark matter in the
solar neighbourhood.  Helmi \etal\ (2001) estimate that there may be
several hundred kinematically cold `dark streams' passing through the
solar neighbourhood.

If CDM is correct in detail, then we have simply failed to detect or to
recognize the many hundreds of missing objects throughout the Local
Group.  For example, the satellites may be dark simply because baryons
were removed long ago through supernova-driven winds (Dekel \& Silk
1986; Mac Low \& Ferrara 1999). In support of this idea, x-ray halos of
groups and clusters are almost always substantially enriched in metals
([Fe/H] $\ge$ -0.5; Renzini 2000; Mushotzky 1999). In fact, we note
that up to 70\% of the mass fraction in metals is likely to reside in
the hot intracluster and intragroup medium (Renzini 2000).  Another
explanation may be that the absence of baryons in hundreds of dark
satellites was set in place long ago during the reionization epoch.
Many authors note that the accretion of gas on to low-mass halos and
subsequent star formation is heavily suppressed in the presence of a
strong photoionizing background (Ikeuchi 1986; Rees 1986; Babul \& Rees
1992; Bullock\etal\ 2000). This effect appears to have a cut-off at low
galactic mass at a characteristic circular velocity close to
30\kms\ (Thoul \& Weinberg 1996; Quinn\etal\ 1996), such that the small
number of visible Local Group dwarfs are those which exceed this cutoff
or acquired most of their neutral hydrogen before the reionization
epoch.

Blitz\etal\ (1999) suggested that the high-velocity \HI\ gas cloud
(HVC) population is associated with dark mini-halos on megaparsec
scales within the Local Group.  This model was refined by Braun \&
Burton (1999) to include only the compact HVCs.  The HVCs have long
been the subject of wide-ranging speculation. Oort (1966) realized that
distances derived from the virial theorem and the HI flux would place 
many clouds at Mpc distances if they are self-gravitating.  If the
clouds lie at about a Mpc and are associated with dark matter clumps,
then they could represent the primordial building blocks.  However
H$\alpha$ distances (Bland-Hawthorn\etal\ 1998) suggest that most HVCs
lie within 50 kpc and are unlikely to be associated with dark matter
halos (Bland-Hawthorn \& Maloney 2001; Weiner\etal\ 2001). We note that
several teams have searched for but failed to detect a faint stellar
population in HVCs.

Moore\etal\ (1999; see also Bland-Hawthorn \& Freeman 2000) suggested
that ultrathin disks in spirals are a challenge to the CDM picture in
that disks are easily heated by orbiting masses.  However, Font\etal\ 
(2001) find that in their CDM simulations, very few of the CDM
sub-halos come close to the optical disk. 

At present there are real problems in reconciling the predictions of
CDM simulations with observations on scales of the Local Group.

\subsection{\bf First order signatures -- information preserved since the
main epoch of baryon dissipation}

\subsubsection{The structure of the disk}

At what stage in the evolution of the disk are its global properties
defined ?  In part, we have already discussed this question in
``Signatures of the internal distribution of specific angular
momentum'' above.  The answer depends on how the internal angular
momentum distribution ${\cal M}(h)$ has evolved as the disk dissipated
and various non-axisymmetric features like bars and spiral structure
came and went. Viscous processes associated with star formation, as
suggested by Lin \& Pringle (1987), may also contribute to the
evolution of the ${\cal M}(h)$ distribution.

The global structure of disks is defined by the central surface
brightnes $I_o$ and the radial scalelength $h$ of the disk. de~Jong \&
Lacey (2000) evaluated the present distribution of galaxies in the
($I_o,h$) plane (Fig.~\ref{fig4}).  If ${\cal M}(h)$ has indeed
remained roughly constant, as is often assumed for discussions of disk
formation (\eg\ Fall \& Efstathiou 1980; Fall 1983), then the global
parameters of the disk $-$ the scale length, central surface brightness
and the Tully-Fisher relation $-$ are relics of the main epoch of
baryon dissipation.

\subsubsection{Can disks preserve fossil information?}

Here, we consider radial and vertical fossil gradients in the disk, in
particular of abundance and age. Our expectation is that much of the
information will be diluted through the dynamical evolution and radial
mixing of the disk. 

For spirals, different mechanisms may be at work to establish gradients
(Molla \etal\ 1996):  (a) a radial variation of the yield due either to
the stellar metal production or to the initial mass function, (b) a
radial variation of the timescale for star formation, (c) a radial
variation for the timescale of infall of gas from outside the disk.
Once gradients are established, these can be amplified or washed out by
radial mixing (Edmunds 1990; Goetz \& Koeppen 1992).

Most stars are born in large clusters numbering hundreds or even
thousands of stars.  Some clusters stay together for billions of years,
whereas others become unbound shortly after the initial starburst,
depending on the star formation efficiency. When a cloud disperses,
each star suffers a random kick superimposed on the cloud's mean
motion.  Thereafter, stars are scattered by transient spiral arm
perturbations and star-cloud encounters.

These perturbations allow the star to migrate in integral space.
During interaction with a single spiral event of pattern speed
$\Omega_p$, a star's energy and angular momentum change while it
conserves its Jacobi integral: in the ($E, J$) plane, stars move along
lines of constant I$_J = E - \Omega_p J$.  The star undertakes a random
walk in the ($E, J$) plane, perturbed by a series of spiral arm events
(Sellwood 1999; Dehnen 2000).  N-body models of disk evolution indicate
that radial mixing is strong (Sellwood 2001; Lynden-Bell \& Kalnajs
1972).  This is believed to be driven by {\it transient} spiral waves
that heat the in-plane motions, although the process is not yet well
understood. Long-term spiral arms produce no net effect. Remarkably, a
{\it single} spiral wave near co-rotation can perturb the angular
momentum of a star by $\sim 20$\% without significant heating: the star
is simply moved from one circular orbit to another, inwards or
outwards, by up to 2~kpc (Sellwood \& Kosowsky 2002). Substantial
variations in the angular momentum of a star are possible over its
lifetime.

In addition to radial heating, stars experience vertical disk heating:
their vertical velocity dispersion increases as they age.  This is
believed to occur through a combination of in-plane spiral-arm heating
and scattering off giant molecular clouds (\eg\ Spitzer \&
Schwarzschild 1953; Carlberg \& Sellwood 1985).  The in-plane heating
is most effective at the inner and outer Lindblad resonances and
vanishes at corotation.  In the vertical direction, an `age-velocity
dispersion' relation is observed for stars younger than about $3$~Ga,
but older disk stars show a velocity dispersion that is independent of
age (Fig.~\ref{fig5}).  Thus, the vertical structure does depend on the 
mean age of the population for $\tau < 3$~Ga (Edvardsson\etal\ 1993, 
confirmed from Hipparcos data by Gomez\etal\ 1997).

As the amplitude of the random motions increases, the star becomes less
vulnerable to heating by transient spiral waves, and the heating
process is expected to saturate. This probably happens after about 3 Ga
(Binney \& Lacey 1988; Jenkins \& Binney 1990), consistent with
observation.  This is important for our purpose here. It means that 
dynamical information is preserved about the state of the thin disk at 
an early epoch, or roughly $\tau_L - 3 \approx 7$ Ga ago, for which 
$\tau_L$ is the look-back time when the disk first began to form.

The survival of old open clusters like NGC 6791, Berkeley 21 and
Berkeley 17 (Friel 1995; van den Bergh 2000) is of interest here.
The oldest open clusters exceed 10~Ga in age and constitute important
fossils (Phelps \& Janes 1996). Both old and young open clusters are part of
the thin disk.  If the heating perturbations occur over a lengthscale
that significantly exceeds the size of an open cluster, it seems likely
that the cluster will survive.  A large spiral-arm heating event will
heat many stars along their I$_J$ trajectories. The trace of the
heating event is likely to survive for a very long time but be visible
only in integral space (Sellwood 2001). We note that vertical
abundance gradients have not been seen among the open clusters (Friel
\& Janes 1993).

About 4\% of disk stars are super metal-rich (SMR) relative to the
Hyades (Castro\etal\ 1997).  SMR stars of intermediate age appear to
have formed a few kpc inside of the Solar circle from enriched gas. The
oldest SMR stars appear to come from the Galactic Center:  their
peculiar kinematics and outward migration may be associated with the
central bar (Carraro \etal\ 1998; Grenon 1999).

In summary,  our expectation is that fossil gradients within the disk
are likely to be weak. This is borne out by observations of both the
stars and the gas (Chiappini\etal\ 2001).

The vertical structure of the disk preserves another fossil $-$ the
`thick disk' $-$ which we discuss in the next section. Like the open
clusters, this component also does not show a vertical abundance
gradient (Gilmore \etal\ 1995). In later sections, we argue
that this may be the most important fossil to have survived the early
stages of galaxy formation.

\subsubsection{Disk heating by accretion: the thick disk}

Heating from discrete accretion events also imposes vertical structure
on the disk (Quinn \& Goodman 1986; Walker \etal\ 1996). Such events
can radically alter the structure of the inner disk and the bulge (see
Fig.~\ref{fig3}(d) for an example), and
are currently believed to have generated the thick disk of the Galaxy.

The galactic thick disk was first recognized by Gilmore \& Reid
(1983).  It includes stars with a wide range of metallicity, from -2.2
$\leq$ [Fe/H] $\leq$ -0.5 (Chiba \& Beers 2000): most of the thick disk
stars are in the more metal-rich end of this range.  The velocity
ellipsoid of the thick disk is observed to be $(\sigma_R,\sigma_{\phi},
\sigma_z) = (46\pm 4, 50\pm 4, 35\pm 3)$\kms\ near the sun, with an
asymmetric drift of about 30\kms.  For comparison, the nearby halo has
a velocity ellipsoid $(\sigma_R,\sigma_{\phi}, \sigma_z) = (141\pm 11,
106\pm 9, 94\pm 8)$\kms\ and its asymmetric drift is about 200 \kms.

The mean age of the thick disk is not known.  From photometric
age-dating of individual stars, the thick disk appears to be as old as
the globular clusters. Indeed, the globular cluster 47 Tuc (age
12.5$\pm$1.5~Ga: Liu \& Chaboyer 2000) is often associated with the 
thick disk.

After Quinn \& Goodman (1986), Walker \etal\ (1996) showed in detail
that a low mass satellite could substantially heat the disk as it sinks
rapidly within the potential well of a galaxy with a `live' halo. The
conversion of satellite orbital energy to disk thermal energy is achieved
through resonant scattering.  Simulations of satellite accretion are 
important for understanding the survival of the thin disk and
the origin of the thick disk.  This is particularly relevant within the
context of CDM.  The satellites which do the damage are those that are dense
enough to survive tidal disruption by the Galaxy. We note that even dwarf 
spheroidals which appear fluffy are in fact rather dense objects dominated 
by their dark matter (J. Kormendy \& K. Freeman, in preparation).

It is fortuitous that the Galaxy has a thick disk, since this is not a
generic phenomenon. The disk structure may be vertically stepped as a
consequence of past discrete accretion events.  The
Edvardsson\etal\ (1993) data (Fig.~\ref{fig5}) appears to show an
abrupt increase in the vertical component of the stellar velocity
dispersion at an age of 10~Ga: see also Str\"omgren (1987).  Freeman
(1991) argued that the age--(velocity dispersion) relation shows three
regimes: stars younger than 3~Ga with $\sigma_z \sim 10$\kms, stars
between 3 and 10~Ga with $\sigma_z \sim 20$\kms, and stars older than
10~Ga with $\sigma_z \sim 40$\kms.  The first regime probably arises
from the disk heating process due to transient spiral arms which we 
described in the previous section.  The last regime is the thick disk, 
presumably excited by an ancient discrete event.  

Can we still identify the disrupting event that lead to the thick disk ?
There is increasing evidence now that the globular cluster $\omega$~Cen is
the stripped core of a dwarf elliptical (see ``Globular clusters'' below).
It is possible that the associated accretion event or an event like it
was the event that triggered the thick disk to form.

In summary, it seems likely that the thick disk may provide a
snap-frozen view of conditions in the disk shortly after the main epoch
of dissipation.  Any low level chemical or age gradients would be of
great interest in the context of dissipation models. In this regard,
Hartkopf \& Yoss (1982) argued for the presence of a vertical abundance
gradient in the thick disk, although Gilmore, Wyse \& Jones (1995) find no
such effect. Because stars of the thick disk spend relatively little
time near the galactic plane, where the spiral arm heating and
scattering by giant molecular clouds is most vigorous, radial mixing
within the thick disk is unlikely to remove all vestiges of a gradient.
If our earlier suggestions are right (see ``Signatures of the internal
distribution of specific angular momentum''), we might expect to see
a different truncation radius for the thick disk compared to the thin
disk.

\subsubsection{Is there an age--metallicity relation ?}

Some fossil information has likely been preserved since the main epoch
of baryon dissipation. The inner stellar bulge is a striking example.
It is characterized by old, metal-rich stars which seems to be at odds
with the classical picture where metals accumulate with time (Tinsley 1980).
However, the dynamical timescales in the inner bulge are very
short compared to the outer disk, and would have allowed for rapid
enrichment at early times. This is consistent with the frequent
occurrence of metal-rich cores of galaxies observed at high redshift
(Hamann \& Ferland 1999). The dynamical complexity of the Galactic
bulge may not allow us to determine the sequence of events that gave
rise to it. We anticipate that this will come about from far-field
cosmology (Ellis\etal\ 2000).

The existence of an age-metallicity relation (AMR) in stars is a very
important issue, about which there has long been disagreement.  Twarog
(1980) and Meusinger\etal\ (1991) provide evidence for the presence of
an AMR, while Carlberg \etal\ (1985) find that the metallicity of
nearby F stars is approximately constant for stars older than about
4~Ga.  More recently it has become clear that an AMR is apparent only
in the solar neighbourhood and is strictly true only for stars younger
than 2~Ga and hotter than $\log\ T_{\rm eff} = 3.8$ (Feltzing\etal\ 2001).
Edvardsson\etal\ (1993) demonstrate that there is no such relation for
field stars in the old disk. Similarly, Friel (1995) shows that there is
no AMR for open clusters (see ``Open clusters''): she goes on to note that

\begin{quote}
Apparently, over the entire age of the disk, at any position in the
disk, the oldest clusters form with compositions as enriched as those
of much younger objects.  
\end{quote}

\noindent In fact, it has been recognized for a long time (\eg\ Arp
1962; Eggen \& Sandage 1969; Hirshfeld\etal\ 1978) that old, metal-rich
stars permeate the galaxy, throughout the disk, the bulge and the halo.
We regard the presence of old metal-rich stars as a first-order
signature. An age--metallicity relation which applies to all stars
would have been an important second-order signature, but we see no
evidence for such a relation, except among the young stars.

\subsubsection{Effects of environment and internal evolution}

Environmental influences are operating on all scales of the hierarchy
and across all stages of our signature classification, so our attempts
to classify signatures are partly artificial. Within CDM, environmental
effects persist throughout the life of the galaxy.

The parameters that govern the evolution of galaxies are among the key
unknowns of modern astrophysics.  Are the dominant influences internal
(\eg\ depth of potential) or external (\eg\ environment) to galaxies ?
We consider here the effects of environment and internal evolution
on the validity of the first-order signatures of galaxy formation 
(\ie\ the properties that may have been conserved since the main epoch of
baryon dissipation).

The well-known `G dwarf problem' indicates that external influences are
important. A simple closed box model of chemical evolution predicts far
too many metal-poor stars in the solar neighbourhood (Tinsley 1980).
This problem is easily remedied by allowing gas to flow into the region
(Lacey \& Fall 1983, 1985; Clayton 1987, 1988; Wyse \& Silk 1989;
Matteuci \& Francois 1989; Worthey \etal\ 1996).  In the context of
CDM, this is believed to arise from the continued accretion of gas-rich
dwarfs (\eg\ Cole\etal\ 1994; Kauffmann \& Charlot 1998).

Environment is clearly a key factor. Early type galaxies are highly
clustered compared to late type galaxies (Hubble \& Humason 1931;
Dressler 1980).  Trager\etal\ (2000) find that for a sample of
early-type galaxies in low-density environments, there is a large
spread in the H$\beta$ index (\ie\ age), but little variation in
metallicity. For galaxies in the Fornax cluster, Kuntschner (2000)
finds the opposite effect:  a large spread in metallicity is present
with little variation in age. This probably reflects strong differences
in environment between the field and the cluster.

Another likely environmental effect is the fraction of S0 galaxies in
clusters, which shows a rising trend with redshift since $z\approx 0.4$
(Jones\etal\ 2000).  Furthermore, S0 galaxies in the Ursa Major cluster
show age gradients that are inverted compared to field spirals, in the
sense that the cores are young and metal-rich (Tully\etal\ 1996;
Kuntschner \& Davies 1998). Both of these effects involve more recent
phenomena and would be properly classified as second-order signatures.

Internal influences are also at work. A manifestation is the
color-magnitude relation (CMR) in early type (Sandage \& Visvanathan
1978) and late type (Peletier \& de Grijs 1998) galaxies.  The CMR does
not arise from dust effects (Bell \& de Jong 2000) and must reflect
systematic variations in age and/or metallicity with luminosity.  In
the case of ellipticals, the CMR is believed to reflect a
mass-metallicity dependence (Faber 1973; Bower \etal\ 1998).  The
relation is naturally explained by supernova-driven wind models in
which more massive galaxies retain supernova ejecta and thus become
more metal rich and redder (Larson 1974; Arimoto \& Yoshii 1987).  The
CMR is presumably established during the main phase of baryon
dissipation and is a genuine first-order signature.

Concannon \etal\ (2000) analyzed a sample of 100 early type galaxies
over a large range in mass. They find that lower mass galaxies exhibit
a larger range in age than higher mass galaxies.  This appears to show
that smaller galaxies have had a more varied star formation history,
which is at odds with the naive CDM picture of low-mass galaxies being
older than high mass galaxies (Baugh \etal\ 1996; Kauffman 1996).  The
work of Concannon\etal\ (2000) shows the presence of a real cosmic
scatter in the star formation history. It is tempting to suggest that
this cosmic scatter relates to different stages of evolution within 
the hierarchy. In this sense, we would regard the Concannon\etal\
result as a first-order manifestation of galaxy formation (see ``Timescales
and fossils'' above).

Spiral galaxies commonly show color gradients that presumably reflect
gradients in age and metallicity (Peletier \& de Grijs 1998).  Faint
spiral galaxies have younger ages and lower metallicities relative 
to bright spirals.  In a study of 120 low-inclination spirals,
Bell \& de Jong (2000) found that the local surface density within 
galaxies is the most important parameter in shaping their star formation
and chemical history. However, they find that metal-rich galaxies occur
over the full range of surface density. This fact has a remarkable
resonance with the distribution of the metal rich open clusters that are
found at any position in the Galactic disk (see ``Is there an age-metallicity
relation?'' above). Bell \& de Jong
argue that the total mass is a secondary factor that modulates the
star formation history.  Once again, these authors demonstrate the
existence of cosmic scatter that may well arise from variations in
environment.

\subsection{\bf Second order signatures -- major processes involved in 
subsequent evolution}

\subsubsection{Introduction}

Here we consider relics of processes that have taken place in the
Galaxy since most of the baryonic mass settled to the disk. There
are several manifestations of these processes, probably the most
significant of which is the star formation history of the disk,
for which the open clusters are particularly important probes.

There is a wealth of detail relating to anomalous populations
throughout the Galaxy, discussed at length by Majewski (1993). 
Examples include an excess of stars on extreme retrograde orbits
(Norris \& Ryan 1989; Carney\etal\ 1996), metal-poor halo stars of
intermediate age (Preston \etal\ 1994) and metal-rich
halo A stars (Rodgers \etal\ 1981).

In an earlier section, we discussed observational signatures of the CDM
hierarchy in the Galactic context. In fact, detailed observations in
velocity space are proving to be particularly useful in identifying
structures that have long since dispersed in configuration space.
In external galaxies, related structures are showing up as low surface
brightness features.  We do not know what role globular clusters play in
the galaxy formation picture, but we include them here because at least
one of them appears now to be the nucleus of a disrupted dwarf galaxy.

\subsubsection{Star formation history}

The star formation history (SFH) of our Galaxy has been very difficult
to unravel. Derived star formation histories range from a roughly
uniform star formation rate over the history of the disk to a SFH that
was highly peaked at early times (\eg\ Twarog 1980; Rocha-Pinto \etal\
2000; Just 2001).  Galaxies of the Local Group show a great diversity
in SFH (Grebel 2001), although the average history over the Local Group
appears consistent with the mean cosmic history (Hopkins\etal\ 2001).
The present emphasis is on star formation studies that make use of the
integrated properties of external galaxies, but it should be noted that
this is necessarily weighted towards the most luminous populations. Key
results for external galaxies are reviewed in ``Effects of environment
and internal evolution'' above.  It was concluded that environmental
effects are very significant in determining the SFH for individual
galaxies.

The conventional approach to the study of chemical evolution in galaxy
disks is to consider the solar neighbourhood a closed box, and to
assume that it is representative of all disks. Simple mathematical
formulations have developed over the past 40 years (van den Bergh 1962;
Schmidt 1963; Pagel \& Patchett 1975; Talbot \& Arnett 1971; Tinsley
1980; Twarog 1980; Pitts \& Tayler 1989). Most observations are
interpreted within this framework.  The SFH is quantified in terms of
stellar age, stellar ($+$ gas) metallicity and, to a lesser extent, the
existing gas fraction.

The use of broadband photometry coupled with stellar population
synthesis is a well-established technique for probing the SFH of galaxy
populations from integrated light. The power of the method is its
simplicity, although it cannot uniquely disentangle the age-metallicity
degeneracy (Bica\etal\ 1990; Charlot \& Silk 1994).

Another widely used technique is the Lick index system (Burstein\etal\
1984) further refined in Worthey\etal\ (1994) and Trager\etal\ (1998).
In this system, the H$\beta$ index is the primary age-sensitive
spectral indicator, whereas the Mg and Fe indices are the primary
metallicity indicators. The Lick indices have well known limitations:
they correspond to low spectroscopic resolution (8$-$9\AA), require
difficult corrections for internal galaxy motions, and are not
calibrated onto a photometric scale. Furthermore, two of the most
prominent Lick indices -- Mg$_2$ $\lambda$5176 and Fe $\lambda$5270 --
are now known to be susceptible to contamination from other elements,
in particular Ca and C (Tripicco \& Bell 1995).

How best to measure galaxy ages is a subject with a long history. The
most reliable methods to date involve the low order transitions
($n<4$) of the Balmer series.  Ages derived from the H$\gamma$
equivalent width have been used by Jones \& Worthey (1995).
Rose (1994) and Caldwell \& Rose (1998) have pioneered the use of even
higher order Balmer lines to break the age-metallicity degeneracy
(Worthey 1994). These higher order lines are less affected by
Balmer line emission from the interstellar medium.  They develop a line
ratio index H$n$/Fe which is a sum over H$\gamma$, H$\delta$ and H8
lines with respect to local Fe lines. The most recent demonstration of
the power of this index can be found at Concannon \etal\ (2000).  

Ultimately, full spectrum fitting matched to spectral synthesis models 
holds the most promise (Vazdekis 1999). The new models, which have a
fourfold increase in spectroscopic resolution compared to the Lick
system, show that the isochrone or isochemical grid lines overlaid on a
plot of two Lick indices are more orthogonal than the Worthey models.
Thus, galaxies like NGC~4365 that exhibit no age gradient in the
Vazdekis models (Davies\etal\ 2001; see Fig.\ref{fig6}(a)) appear to
show an age spread in the Worthey models. Interestingly, NGC~4150 
exhibits an abundance spread with constant age (Fig.\ref{fig6}(b)).

\subsubsection{Low surface brightness structures in galaxies}

Dynamical interaction between galaxies lead to a range
of structures including stellar shells (Malin \& Carter 1980; Quinn 1984), 
fans (Weil \etal\ 1997), and tidal streamers 
(Gregg \& West 1998; Calcaneo-Roldan\etal\ 2000; Zheng\etal\ 1999).
Some excellent examples are shown in Fig.~\ref{fig7}.
We see evidence of multiple nuclei, counter-rotating cores, and gas in
polar orbits. At low light levels, the outermost stellar contours of
spiral disks appear frequently to exhibit departures from axisymmetry
(Rix \& Zaritsky 1995). The same is true for spiral arms in all Hubble
types (Schoenmakers \etal\ 1997; Cianci 2002).

The stellar streamers are particularly interesting as these may
provide important constraints on galaxy models, particularly as kinematic
measurements become possible through the detection of planetary
nebulae.  More than a dozen stellar streams are already known and this
is probably indicative of a much larger population at very low surface
brightness. Johnston \etal\ (2001) show that stellar
streamers can survive for several gigayears and are only visible above
the present optical detection limit ($\mu_V \sim 30$ mag arcsec$^{-2}$)
for roughly $4\times 10^8$ yr.  A few galaxy groups (\eg\ the Leo
group) do show largescale HI filaments that can remain visible for many
Ga.

Deep CCD imaging has revealed a stellar loop around NGC 5907
(Shang\etal\ 1998) and a stellar feature extending from NGC 5548
(Tyson\etal\ 1998). The technique of photographic amplification has
revealed stellar streamers in about ten sources (Malin \& Hadley 1997;
Calcaneo-Roldan\etal\ 2000; Weil\etal\ 1997). For these particular
observations, the limiting surface brightness is $\mu_V \approx 28.5$
mag arcsec$^{-2}$. For all of these systems, we estimate that the total
stream luminosities are in the range $3-20\times 10^7$L$_\odot$.

In a recent development, wide-field CCD cameras have revealed stellar
streamers through multiband photometry of millions of individual
sources. A pointillist image can then be reconstructed 
in narrow color intervals so as to enhance features with respect to
the field.  This has lead to the discovery of a stellar stream in M31
(Ibata\etal\ 2001$a$) and tidal tails extending from the globular cluster
Pal 5 (Odenkirchen\etal\ 2001). This technique has the potential to
push much deeper than the direct imaging method described above. 

The low surface brightness universe is notoriously difficult to
observe.  Modern telescope and instrument designs are simply not
optimized for this part of parameter space. Many claims of diffuse
light detections in the neighbourhood of galaxies have been shown to
arise from scattered light internal to the instrument.

In ``Structures in phase space,'' we discuss `moving groups' identified
within the Galaxy from proper motion and spectroscopic surveys. Their
projected surface brightness is $\mu_V = 30-34$ mag arcsec$^{-2}$,
below the limit of modern imaging techniques.

Looking farther afield, we see evidence for discrete accretion events in
the making. The Galaxy is encircled by satellite galaxies that appear
confined to one or two great streams across the sky (Lynden-Bell \& 
Lynden-Bell 1995).  The most
renowned of these are the Magellanic Clouds and the associated HI
Magellanic stream. All of these are expected to merge with the Galaxy in
the distant future, largely due to the dynamical friction from the extended
halo.

\subsubsection{Open clusters }  

In the context of near-field cosmology, we believe that the thick
disk and the old open clusters of the thin disk are among the most 
important diagnostics. The open clusters are the subject of an
outstanding and comprehensive review by Friel (1995). Here, we
summarize the properties that are most important for our purpose.

Both old and young clusters are part of the thin disk.  Their key
attribute is that they provide a direct time line for investigating
change, which we explore in ``The Gaiasphere and the limits of
knowledge.''  The oldest open clusters exceed 10~Ga in age and
constitute important fossils (Phelps \& Janes 1996). In ``Can disks
preserve fossil information?'', we noted that the survival of these
fossil clusters is an interesting issue in its own right.  Friel (1995)
finds no old open clusters within a galactocentric radius of 7~kpc;
these are likely to have disrupted or migrated out of the central
regions (van den Bergh \& McClure 1980). It has long been recognized
that open clusters walk a knife edge between survival and disruption
(King 1958$a,b,c$).

Like field stars in the disk, Janes \& Phelps (1994) find that the old
cluster population (relative to Hyades) is defined by a 375~pc scale height
exponential distribution, whereas young clusters have a 55~pc scale
height (Fig.~\ref{fig8}(a),(b)). Again, like the field stars,
vertical abundance gradients have not been seen in open clusters
(Friel \& Janes 1993), although radial gradients are well established
(Friel 1995; van den Bergh 2000). For old open clusters, Twarog\etal\
(1997) claim evidence for a stepped radial metallicity distribution where
[Fe/H]$\approx$0 within 10~kpc, falling to [Fe/H]$\approx$-0.3 in the
outer disk.  However, this effect is not seen in young objects, \eg\
\HII\ regions and B stars (Henry 1998).

In Fig.~\ref{fig8}(c), both the old and young open clusters show
essentially the same radial trend in metallicity. After reviewing the
available observations, Friel (1995) finds no evidence for an
age-metallicity relation for open clusters (Fig.~\ref{fig8}(d)). In
agreement with Eggen \& Sandage (1969), she notes that over the entire
age of the disk, at any position in the disk, the oldest clusters form
with compositions as enriched as those of much younger objects.

These remarkable observations appear to indicate that shortly after the
main epoch of baryon dissipation, the thin disk was established at
least as far out as 15 kpc.  The oldest open clusters approach the age
of the thick disk.  Since, in ``Disk heating by accretion,'' we noted
that the thick disk is likely to be a `snap frozen' picture of the thin
disk shortly after disk formation, we would expect the truncation of the
thick disk (see ``Signatures of global quantities'') to reflect the extent
of the thin disk at the epoch of the event that puffed up the thick disk.

\subsubsection{Globular clusters} 

We have long suspected that globular clusters are the fossil remnants
of violent processes in the protogalactic era (Peebles \& Dicke 1968).
But there is a growing suspicion that globulars are telling us more
about globulars than galactic origins (Harris 2001).  The Milky Way has
about 150 globular clusters with 20\% lying within a few kiloparsecs of
the Galactic Center. They constitute a negligible fraction of the light
and mass ($2\%$) of the stellar halo today. Their significance rests in
their age.  The oldest globular clusters in the outer halo have an age
of 13 $\pm$ 2.5 Ga (90\% confidence).

The ages of the oldest globular clusters in the inner and outer halo,
the Large Magellanic Cloud and the nearby Fornax and Sgr dwarf
spheroidal galaxies show a remarkable uniformity. To a precision of
$\pm 1$ Ga, the onset of globular cluster formation was well
synchronized over a volume centered on our Galaxy with a radius $>
100$ kpc (Da Costa 1999).

Globular cluster stars are older than the oldest disk stars, \eg\ white
dwarfs and the oldest red giants.  These clusters are also more metal
poor than the underlying halo light in all galaxies and at all radii
(Harris 1991), but again there are exceptions to the rule.  Since
Morgan's (1950) and Kinman's (1959) classic work, we have known that
there are two distinct populations of globular clusters in the Galaxy.
The properties that we associate with these two populations today were
derived by Zinn (1985) who showed that they have very different
structure, kinematics and metallicities.  The halo population is
metal-poor ([Fe/H]$<$-0.8) and slowly rotating with a roughly spherical
distribution;  the disk population is metal-rich ([Fe/H]$>$-0.8) and in
rapid rotation.

A major development has been the discovery of young
globular clusters in disturbed or interacting galaxies, e.g., NGC 1275
(Holtzman\etal\ 1992), NGC 7252 (Whitmore\etal\ 1993) and the Antennae 
(Whitmore \& Schweizer 1995).  Schweizer (1987) first
suspected that globular clusters were formed in mergers. Later, Ashman
\& Zepf (1992) predicted that the HST would reveal young globular
clusters through their compact sizes, high luminosities and blue
colors.  The very high internal densities of globular clusters today
must partly reflect the conditions when they were formed.  Harris \&
Pudritz (1994) present a model for globular clusters produced in
fragmenting giant molecular clouds, which are of the right mass and
density range to resemble accretion fragments in the Searle-Zinn
model.

Globular clusters have been elegantly referred to as `canaries in a
coal mine' (Arras \& Wasserman 1999). They are subject to a range of
disruptive effects, including two-body relaxation and erosion by the
tidal field of their host galaxy, and the tidal shocking that they
experience as their orbits take them through the galactic disk and
substructure in the dark halo.  In addition to self-destruction through
stellar mass loss, tidal shocking may have been very important in the
early universe (Gnedin \etal\ 1999).  If globular clusters originally
formed in great numbers, the disrupted clusters may now contribute to
the stellar halo (Norris \& Ryan 1989; Oort 1965).  Halo field stars
and globular clusters in the Milky Way have similar mean metallicities
(Carney 1993); however the metallicity distribution of the halo field
stars extends to much lower metallicity ([Fe/H] $\simeq -5$) than that
of the globular clusters ([Fe/H] $\simeq -2.2$).  We note again the
remarkable similarity in the metallicity range of the globular clusters
and the thick disk ($-2.2 \lta $[Fe/H]$ \lta -0.5$).

In the nucleated dwarf elliptical galaxies (Binggeli \etal\ 1985), the
nucleus typically provides about 1\% of the total luminosity; globular
clusters could be considered as the stripped nuclei of these satellite
objects without exceeding the visible halo mass (Zinnecker \& Cannon
1986; Freeman 1993).  It is an intriguing prospect that the existing
globular clusters could be the stripped relicts of an ancient swarm of
protogalactic stellar fragments, \ie\ the original building blocks of
the Universe.

In the Searle-Zinn picture, globular clusters are intimately linked to
{\it gas-rich}, protogalactic infalling fragments. Multiple stellar
populations have recently been detected in $\omega$~Cen, the most
massive cluster in the Galaxy (Lee\etal\ 1999).  How did $\omega$ Cen
retain its gas for a later burst?  It now appears that it was
associated with a gas-rich dwarf, either as an {\it in situ} cluster or
as the stellar nucleus. The present-day cluster density is sufficiently
high that it would have survived tidal disruption by the Galaxy, unlike
the more diffuse envelope of this dwarf galaxy. The very bound 
retrograde orbit supports the view that $\omega$~Cen entered the Galaxy
as part of a more massive system whose orbit decayed through dynamical
friction.

If globular clusters are so ancient, why are the abundances of the most
metal-poor population as high as they are? Because it does not take
much star formation to increase the metal abundance up to [Fe/H]$=$-1.5
(Frayer \& Brown 1997), the cluster abundances may reflect low levels
of star formation even before the first (dark$+$baryon) systems came
together. 

Old age is not necessarily associated with low metallicity (cf.
``Timescales and fossil'' above).  We recall that CO has been detected
at $z\sim 5$ (Yun\etal\ 2000).  Hamann \& Ferland (1999) demonstrate
that stellar populations at the highest redshift currently observed
appear to have solar or super-solar metallicity. We believe that there
is no mystery about high abundances at high redshift.  The dynamical
times in the cores of these systems are short, so there has been time
for multiple generations of star formation and chemical enrichment.
In this sense, the cores of high redshift galaxies need not be relevant
to the chemical properties of the globular clusters, although both kinds
of objects were probably formed at about the same time.

The first generation of globular clusters may have been produced in
merger-driven starbursts when the primordial fragments came together
for the first time.  If at least some fragments retained some of their
identity while the halo was formed, a small number of enrichment events
per fragment would ensure a Poissonian scatter in properties between
globular clusters, and multiple populations within individual clusters
(Searle \& Zinn 1978).  

\subsubsection{Structures in phase space}

One class of systems that exhibit coherence in velocity space are the
open clusters associated with the disk.  Here the common space motion
of the stars with respect to the Sun is perceived as a convergence of
the proper motions to a single point (strictly speaking, minimum
volume) on the sky (Boss 1908; see de Zeeuw\etal\ 1999 for a recent
application).  More than a dozen such
systems have been identified this way. However, these are all young
open clusters largely associated with the Gould belt.  With
sufficiently precise kinematics, it may be possible to identify open
clusters that have recently dispersed, particularly if the group is
confined to a specific radial zone by resonances in the outer disk.
For example, Feltzing \& Holmberg (2000) show that the metal-rich
([Fe/H]$\approx 0.2$) moving group HR~1614, thought to be 2~Ga old, can
be identified in the Hipparcos data set.

Recently, attention has turned to a diverse set of `moving groups'
that are thought to be associated with the stellar halo and in some
instances are clearly fossils associated with accretion events in the
distant past. The evidence for these groups dates back to the
discovery of the halo itself.  Shortly before the publication of the
landmark ELS paper, Eggen \& Sandage (1959) discovered that the nearby
high-velocity star, Groombridge 1830, belongs to a moving group now
passing through the Galactic disk.

In a long series of papers, Eggen went on to `identify' a number of
moving groups, some of which appear to encompass the solar
neighbourhood, and others that may be associated with the halo. The
relevant references are given by Taylor (2000). Various authors have
noted that many of the groups are difficult to confirm (Griffin 1998;
Taylor 2000).  More systematic surveys over the past few decades have
identified a number of moving populations associated with the halo
(Freeman 1987; Majewksi 1993) although the reality of some of these
groups is still debated. The reality of these groups is of paramount
importance in the context of halo formation.  Majewski\etal\ (1996)
suspect that much or all of the halo could exhibit phase-space clumping
with data of sufficient quality.

In recent years, the existence of kinematic sub-structure in the
galactic halo has become clear. Helmi\etal\ (1999) identified 88
metal-poor stars within 1 kiloparsec of the Sun from the Hipparcos
astrometric catalogue. After deducing accurate 3-D space motions, they
found a highly significant group of 8 stars that appear clumped in
phase space and confined to a highly inclined orbit.

The most dramatic evidence is surely the highly disrupted Sgr dwarf
galaxy identified by Ibata\etal\ (1994; 1995). These authors used
multi-object spectroscopy to uncover an elongated stellar stream moving
through the plane on the far side of the Galaxy.  The Sgr
dwarf is a low mass dwarf spheroidal galaxy about 25 kpc from the Sun
that is presently being disrupted by the Galactic tidal field. The
long axis of the prolate body (axis ratios $\sim$ 3:1:1) is about 10
kpc, oriented perpendicular to the Galactic plane along $\ell=6^\circ$
and centered at $b=-15^\circ$. Sgr contains a mix of stellar
populations, an extended dark halo (mass $\geq$ 10$^9$M$_\odot$) and at
least four globular clusters (Ibata\etal\ 1997).  The Sgr stream has
since been recovered by several photometric surveys (Vivas\etal\ 2001; 
Newberg \etal\ 2002; Ibata \etal\ 2001$c$).

N-body simulations have shown that stellar streams are formed when low
mass systems are accreted by a large galaxy (\eg\ Harding\etal\ 2001).
Streamers remain dynamically cold and identifiable as a kinematic
substructure long after they have ceased to be recognizable in star
counts against the vast stellar background of the galaxy (Tremaine
1993; Ibata \& Lewis 1998; Johnston 1998; Helmi \& White 1999).

Within the Galaxy, moving groups can be identified with even limited
phase-space information (de~Bruijne 1999; de~Zeeuw\etal\ 1999). This
also holds for satellites orbiting within the spherical halo, since the
debris remains in the plane of motion for at least a few orbits
(Lynden-Bell \& Lynden-Bell 1995; Johnston \etal\ 1996).
But a satellite experiencing the disk potential no longer conserves its
angular momentum and its orbit plane undergoes strong precession (Helmi
\& White 1999).  In Fig.~\ref{fig9}, we show the sky projection of a
satellite 8~Ga after disruption. These more complex structures are
usually highly localized, and therefore easy to recognize in the space
of conserved quantities like energy and angular momentum for individual
stars.

The evolution in phase space of a disrupting satellite is well behaved
as its stars become phase mixed. Its phase space flow obeys Liouville's
theorem, \ie\ the flow is incompressible. Highly intuitive accounts are
given elsewhere (Carlberg 1986; Tremaine 1999; Hernquist \& Quinn
1988).  It should be possible to recognize partially phase-mixed
structures that cover the observed space, although special techniques
are needed to find them.

Four astrometric space missions are planned for the next decade.
These are the proposed German DIVA mission ($\sim 2003$); the
FAME mission ($\sim 2005$) and the pointed SIM mission ($\sim 2005$); 
and the ESA Gaia mission ($\sim 2009$) which will observe a billion
stars to V$\sim$20, with accuracy $10 \mu$as at V$\sim$15.
The web sites for these missions are at:
http://www.ari.uni-heidelberg.de/diva/,
http://aa.usno.navy.mil/FAME/,
http://sim.jpl.nasa.gov/,
http://astro.estec.esa.nl/GAIA/.

The astrometric missions will derive 6-dimensional phase space
coordinates and spectrophotometric properties for millions of stars
within a 20 kiloparsec sphere $-$ the Gaiasphere. The ambitious Gaia
mission will obtain distances for up to 90 million stars with better
than 5\% accuracy, and measure proper motions with an accuracy
approaching {\it micro}arcsec per year.  If hierarchical CDM is
correct, there should be thousands of coherent streamers that make up
the outer halo, and hundreds of partially phase-mixed structures within
the inner halo.  A satellite experiencing the disk potential no longer
conserves its angular momentum and its orbit plane undergoes strong
precession (see Fig.~\ref{fig10}(c) and (d)).  In Fig.~\ref{fig10}(a)
and (b), Helmi \etal\ (1999) demonstrate the relative ease with which
Gaia will identify substructure within the stellar halo.

\section{The Gaiasphere and the limits of knowledge}

\subsection{Introduction}

The ultimate goal of cosmology, both near and far, must be to explain
how the Universe has arrived at its present state. It is plausible
$-$ although difficult to accept $-$ that nature provides fundamental 
limits of knowledge, in particular, epochs where the sequence of events 
are scrambled. Our `intuition' is that any phase dominated by relaxation 
or dissipation probably removes more information than it retains.

But could some of the residual inhomogeneities from prehistory have
escaped the dissipative process at an early stage? We may not know the
answer to this question with absolute certainty for many years. In the
absence of certainty, we consider what might be the likely traces of a
bygone era prior to the main epoch of baryon dissipation.

\subsection{Chemical signatures}

A major goal of near-field cosmology is to tag or to associate individual
stars with elements of the protocloud.  For many halo stars, and some
outer bulge stars, this may be possible with phase space information
provided by Gaia. But for much of the bulge and the disk, secular
processes cause the populations to become relaxed (\ie\ the integrals
of motion are partially randomized).  In order to have any chance of
unravelling disk formation, we must explore chemical signatures in the
stellar spectrum.  Ideally, we would like to tag a large sample of
representative stars with a precise {\it time} and a precise {\it site}
of formation.

Over the last four decades, evidence has gradually accumulated
(Fig.~\ref{fig11}) for a large dispersion in metal abundances [X$_i$/Fe]
(particularly n-capture elements) in low metallicity stars relative to
solar abundances (Wallerstein\etal\ 1963; Pagel 1965; Spite \& Spite
1978; Truran 1981; Luck \& Bond 1985; Clayton 1988; Gilroy\etal\ 1988;
McWilliam\etal\ 1995; Norris \etal\ 1996; Burris\etal\ 2000).  Elements
like Sr, Ba and Eu show a 300-fold dispersion, although [$\alpha$/Fe]
dispersions are typically an order of magnitude smaller.

In their celebrated paper, Burbidge\etal\ (1957 $-$ B$^2$FH)
demonstrated the likely sites for the synthesis of slow (s) and rapid
(r) n-capture elements. The s-process elements (\eg\ Sr, Zr, Ba, Ce,
La, Pb) are thought to arise from the He-burning phase of
intermediate to low mass (AGB) stars (M $<$ 10M$_\odot$), although
at the lowest metallicities, trace amounts are likely to arise from
high mass stars (Burris\etal\ 2000; Rauscher\etal\ 2001).

In contrast to the s-process elements, the r-process elements (\eg\ Sm,
Eu, Gd, Tb, Dy, Ho) cannot be formed during quiescent stellar
evolution.  While some doubts remain, the most likely site for the
r-process appears to be SN~II, as originally suggested by B$^2$FH (see
also Wallerstein\etal\ 1997). Therefore, r-process elements measured
from stellar atmospheres reflect conditions in the progenitor cloud. In
support of Gilroy\etal\ (1988), McWilliam\etal\ (1995) state that
`the very large scatter means that n-capture element abundances in
ultra-metal poor stars are products of one or very few prior
nucleosynthesis events that occurred in the very early, poorly mixed
galactic halo', a theme that has been developed by many authors
(\eg\ Audouze \& Silk 1995; Shigeyama \& Tsujimoto 1998; Argast\etal\
2000; Tsujimoto \etal\ 2000).

Supernova models produce different yields as a function
of progenitor mass, progenitor metallicity, mass cut (what gets
ejected compared to what falls back towards the compact central
object), and detonation details. The $\alpha$ elements are mainly 
produced in the hydrostatic burning phase within the pre-supernova star.  
Thus $\alpha$ yields are not dependent on the mass cut or
details of the fallback/explosion mechanism which leads to a much
smaller dispersion at low metallicity.

There is no known age-metallicity relation that operates over a useful
dynamic range in age and/or metallicity. (This effect is only seen in a
small subset of hot metal-rich stars -- see ``Is there an
age-metallicity relation?'' above).  Such a relation would require the
metals to be well mixed over large volumes of the ISM.  For the
forseeable future, it seems that only a small fraction of stars can be
dated directly (see ``Stellar age dating'' above).

\subsection{Reconstructing ancient star groups}

We now conjecture that the heavy element metallicity dispersion may
provide a way forward for tagging groups of stars to common sites of
formation.  With sufficiently detailed spectral line information, it is
feasible that the `chemical tagging' will allow temporal sequencing of
a large fraction of stars in a manner analogous to building a family
tree through DNA sequencing.

Most stars are born within rich clusters of many hundreds to many
thousands of stars (Clarke\etal\ 2000; Carpenter 2000).  McKee \& Tan
(2002) propose that high-mass stars form in the cores of strongly
self-gravitating and turbulent gas clouds. The low mass stars form
within the cloud outside the core, presumably at about the same time or
shortly after the high mass stars have formed. The precise sequence of
events which give rise to a high mass star is a topic of great interest
and heated debate in contemporary astrophysics
(\eg\ Stahler\etal\ 2000).

A necessary condition for `chemical tagging' is that the progenitor
cloud is uniformly mixed in key chemical elements before the first stars
are formed.  Another possibility is that a few high mass stars form
shortly after the cloud assembles, and enrich the cloud fairly uniformly.
Both scenarios would help to ensure that long-lived stars have identical
abundances in certain key elements before the onset of low-mass star
formation.

For either statement to be true, an important requirement is that {\it open
clusters of any age have essentially zero dispersion in some key metals
with respect to Fe}.  There has been very little work on heavy element
abundances in open clusters. The target clusters must have reliable
astrometry so as to minimize `pollution' from stars not associated with
the cluster (Quillen 2002).

If our requirement is found not to be true, then either the progenitor
clouds are not well mixed or high mass stars are formed after most
low mass stars. A more fundamental consequence is that {\it a direct
unravelling of the disk into its constituent star groups would be
impossible, in other words, the epoch of dissipation cannot be unravelled
after all.}

Consider the (extraordinary) possibility that we {\it could} put
many coeval star groups back together over the entire age of the Galaxy.
This would provide an accurate age for the star groups either through
the color-magnitude diagram, or through association with those stars
within each group that have [n-capture/Fe] $\gg$ 0, and can therefore
be radioactively dated.  This would provide key information on the
chemical evolution history for each of the main components of the
Galaxy.

But what about the formation site? The kinematic signatures will
identify which component of the Galaxy the reconstructed star group
belongs to, but not specifically where in the Galactic component
(\eg\ radius) the star group came into existence. For stars in the thin
disk and bulge, the stellar kinematics will have been much affected by
the bar and spiral waves; it will no longer be possible to estimate
their birthplace from their kinematics.  Our expectation is that the
derived family tree will severely restrict the possible scenarios
involved in the dissipation process. In this respect, a sufficiently
detailed model may be able to locate each star group within the
simulated time sequence.

In addition to open clusters, we have already argued that the thick
disk is an extremely important fossil of the processes behind disk
formation. The thick disk is thought to be a snap-frozen relic of the
early disk, heated vertically by the the infall of an intermediate mass
satellite. Chemical tagging of stars that make up the thick disk would
provide clues on the formation of the first star clusters in the early
disk.

\subsection{Chemical abundance space}

An intriguing prospect is that reconstructed star clusters can be
placed into an evolutionary sequence, \ie\ a family tree, based on
their chemical signatures. Let us suppose that a star cluster has
accurate chemical abundances determined for a large number $n$ of
elements (including isotopes). This gives it a unique location in an
$n$-dimensional space compared to $m$ other star clusters within that
space. We write the chemical abundance space as
${\cal C}$(Fe/H, X$_1$/Fe, X$_2$/Fe, ...) where X$_1$, X$_2$ ... 
are the independent chemical elements that define the space (\ie\
elements whose abundances are not rigidly coupled to other elements).

The size of $n$ is unlikely to exceed about 50 for the foreseeable
future. Hill\etal\ (2002) present exquisite data for the metal-poor
star CS~31082-001, where abundance estimates are obtained for a total
of 44 elements, almost half the entire periodic table (see also
Cayrel\etal\ 2001).  In Fig.\ref{fig12}, we show what is now possible
for another metal-poor star, CS~22892-052 (Sneden\etal\ 2001$a$).  The
$\alpha$ elements and r-process elements, and maybe a few canonical
s-process elements at low [Fe/H], provide information on the cloud
abundances prior to star formation, although combinations of these are
likely to be coupled (Heger \& Woosley 2001; Sneden\etal\ 2001$a$).
There are 24 r-process elements that have been clearly identified in
stellar spectra (Wallerstein\etal\ 1997).

The size of $m$ is likely to be exceedingly large for the thin disk
where most of the baryons reside. For a rough estimate, we take the age
of the disk to be 10~Ga. If there is a unique SN~II enrichment event
every 100 years, we expect of order 10$^8$ formation sites.  Typically,
a SN~II event sweeps up a constant mass of $5\times 10^4$M$_\odot$
(Ryan\etal\ 1996; Shigeyama \& Tsujimoto 1998). Simple chemical
evolution models indicate that this must be of the right order to
explain the metallicity dispersion at low [Fe/H] (Argast\etal\ 2000).
Roughly speaking, there have been 10$^3$ generations of clouds since
the disk formed, with about 10$^5$ clouds in each star-forming
generation, such that cloud formation and dispersal cycle on a 10$^7$
yr timescale (Elmegreen\etal\ 2000).

Whereas the total number of star clusters over the lifetime of the thin
disk is very large, the size of $m$ for the stellar halo
(Harding\etal\ 2001), and maybe the thick disk (Kroupa 2002), is likely
to be significantly smaller.  Our primary interest is the oldest star
clusters.  Reconstructing star clusters within the thick disk is a
particularly interesting prospect since the disk is likely to have
formed within 1$-$1.5~Ga of the main epoch of baryon dissipation
(Prochaska\etal\ 2000).

The task of establishing up to 10$^8$ unique chemical signatures may
appear to be a hopeless proposition with current technology. But it
is worth noting that more than 60 of the chemical elements ($Z > 30$)
arise from n-capture processes. Let us suppose that half of these are
detectable for a given star. We would only need to be able to 
measure two distinct abundances for each of these elements in 
order to achieve 10$^9$ independent cells in ${\cal C}$-space.
If many of the element abundances are found to be {\it rigidly}
coupled, of course the parameter space would be much smaller.

It may not be necessary to measure as many as 30 elements if some
can be found which are highly decoupled and exhibit large relative
dispersions from star to star. Burris\etal\ (2000) demonstrate one 
such element pair, i.e. [Ba/Fe] and [Sr/Fe]. It is difficult at 
this stage to suggest which elements are most suited to chemical
tagging. In part, this depends on the precise details and mechanism
of formation of the n-capture elements at low [Fe/H]. 

The element abundances [X$_i$/Fe] show three main peaks at Z $\approx$
26, Z $\approx$ 52, and Z $\approx$ 78; the last two peaks are evident
in Fig.~\ref{fig12}. There have been
suggestions that the r-process gives rise to random abundance patterns
(\eg\ Goriely \& Arnould 1996) although this is not supported by new
observations of a few metal poor stars.  Heavy r-process elements
around the second peak compared to the Sun appear to show a universal
abundance pattern (Sneden\etal\ 2000; Cayrel\etal\ 2001;
Hill\etal\ 2002). However, Hill\etal\ find that the third peak and
actinide elements (Z $\geq$90) are decoupled from elements in the second
peak. We suspect that there may be a substantial number of suitable
elements ($\sim$10) which could define a sufficiently large parameter
space. 

Our ability to detect structure in ${\cal C}$-space depends on how
precisely we can measure abundance differences between stars. It may be
possible to construct a large database of differential abundances from
echelle spectra, with a precision of 0.05 dex or better; differential
abundances are preferred here to reduce the effects of systematic
error.

\subsection{Chemical trajectories}

Our simple picture assumes that a cloud forms with a unique chemical
signature, or that shortly after the cloud collapses, one or two
massive SN II enrich the cloud with unique yields which add to the
existing chemical signature. The low-mass population forms with this
unique chemical signature. If the star-formation efficiency is high
($\gta 30$\%), the star group stays bound although the remaining gas is
blown away. If the star-formation efficiency is low ($\lta 10$\%), the
star cluster disperses along with the gas. In a closed box model, the
dispersed gas reforms a cloud at a later stage.

In the closed box model, each successive generation of supernovae
produces stellar populations with progressive enrichments. These will
lie along a trajectory in ${\cal C}$-space which can be identified in
principle using minimum spanning tree methods (Sedgewick 1992).  The
overall distribution of the trajectories will be affected by
fundamental processes like the star formation efficiency, the star
formation timescale, the mixing efficiency, the mixing timescale, and
the satellite galaxy infall rate.

As we approach solar levels of metallicity in [Fe/H], the vast number 
of trajectories will converge. By [Fe/H] $\approx$ -2.5, AGB stars will
have substantially raised the s-process element abundances; by [Fe/H]
$\approx$ -1, Type Ia supernovae will have raised the Fe-group
abundances.  Star clusters that appear to originate at the same
location in this ${\cal C}$-space may simply reflect a common formation
site, \ie\ the resolution limit we can expect to achieve in
configuration space. The ability to identify common formation sites
rests on accurate differential abundance analyses
(Edvardsson\etal\ 1993; Prochaska\etal\ 2000).

Even with a well established family tree based on chemical trajectories
in the chemical ${\cal C}$-space, this information may not give a clear
indication of the original location within the protocloud or Galactic
component. This will come in the future from realistic baryon
dissipation models. Forward evolution of any proposed model must be able
to produce the chemical tree.

However, the ${\cal C}$-space will provide a vast amount of
information on chemical evolution history. It should be possible to
detect the evolution of the cluster mass function with cosmic time
(Kroupa 2002), the epoch of a starburst phase and/or associated mass
ejection of metals to the halo (Renzini 2001), and/or satellite infall
(Noguchi 1998).

As we go back in time to the formation of the disk, we approach the
chemical state laid down by population~III stars.  The lack of stars
below [Fe/H] $\approx$ -5 suggests that the protocloud was initially
enriched by the first generation of stars (Argast\etal\ 2000). However,
the apparent absence of any remnants of population~III remains a
puzzle: its stars may have had a top-heavy initial mass function, or
have dispersed into the intra-group medium of the Local Group.  If one
could unravel the abundances of heavy elements at the time of disk
formation, this would greatly improve the precision of
nucleo-cosmochronology (see ``Stellar age dating'').

\subsection{Candidates for chemical tagging}

Chemical tagging will not be possible for all stars.  In hot stars, 
our ability to measure abundances is reduced by the stellar rotation
and lack of transitions for many ions in the optical.
The ideal candidates are the evolved FGK stars that are numerous
and intrinsically bright. These can be observed at echelle
resolutions ($R > 30,000$) over the full Gaiasphere. Moreover, giants
have deep, low density atmospheres that produce strong low-ionization
absorption lines compared to higher gravity atmospheres.  Even in the
presence of significant line blending, with sufficient signal, it
should be possible to derive abundance information by comparing the
fine structure information with accurate stellar synthesis models.
Detailed abundances of large numbers of F and G subgiants would be
particularly useful, if it becomes possible to make such studies,
because direct relative ages can be derived for these stars from their
observed luminosities.

It is not clear at what [Fe/H] the r-process elements become swamped by
the ubiquitous Fe-group and s-process elements.  At a resolution of
$R\sim 10^5$, many r-process elements can be seen in the solar
spectrum, although the signal-to-noise ratio of about 1000 is needed,
and even then the spectral lines are often badly blended (Kurucz 1991;
1995).  Travaglio\etal\ (1999) suggest that the s-process does not
become significant until [Fe/H] $\approx$ -1 because of the need for
pre-existing seed nuclei (Spite \& Spite 1978; Truran 1981), although
Pagel \& Tautvaisiene (1997) argue for some s-process production at
[Fe/H] $\sim$ -2.5.  Prochaska\etal\ (2000) detected Ba, Y and Eu in a
snapshot survey of thick disk G dwarfs in the solar neighbourhood with
-1.1 $\lta$ [Fe/H] $\lta$ -0.5. This survey only managed to detect a
few transitions in each element although their spectral coverage was
redward of 440nm with SNR $\approx$ 100 per pixel at $R\simeq 50,000$.
Longer exposures with $R\sim 10^5$ and spectral coverage down to 300nm
would have detected more heavy elements.

\subsection{Summary}

In our view, observations of nucleosynthetic signatures of metal-poor
stars provide a cornerstone of near-field cosmology. Success in this
arena requires major progress across a wide front, including better
atomic parameters (Truran\etal\ 2001), improved supernova models,
better stellar synthesis codes and more realistic galaxy formation
models.  There are no stellar evolutionary models that lead to a
self-consistent detonation and deflagration in a core-collapse
supernova event or, for that matter, detonation in a thermonuclear
explosive event.  Realistic chemical production at the onset of the
supernova stage requires a proper accounting of a large number of
isotope networks (400$-$2500) that cannot be adequately simulated
yet.  Modern computers have only recently conquered relatively
simple $\alpha$ networks involving 13 isotopes. The inexorable march of
computer power will greatly assist here.  

There is also a key experimental front both in terms of laboratory
simulations of nucleosynthesis, and the need for major developments in
astronomical instrumentation (see ``Epilogue: challenges for the
future''). Many authors (\eg\ Sneden\etal\ 2001$b$) have stressed the 
importance of greatly improving the accuracy of transition
probabilities and reaction rates for both heavy and light ion
interactions. This will be possible with the new generation of
high-intensity accelerators and radioactive-beam instruments
(K\"{a}ppeler\etal\ 1998; \qv Manuel 2000).

Progress on all fronts will require iteration between the different
strands. Already, relative r-process and $\alpha$ element abundances
for metal-poor stars have begun to constrain the yields for different
stellar masses and associated mass cuts of progenitor supernovae
(Mathews\etal\ 1992; Travaglio\etal\ 1998; Ishimaru \& Wanajo
2000).

\smallskip
It is an intriguing thought that one day we may be able to identify
hundreds or thousands of stars throughout the Gaiasphere that were 
born within the same cloud as the Sun.

\section{EPILOGUE: CHALLENGES FOR THE FUTURE}

Throughout this review, we have identified fossil signatures of galaxy
formation and evolution which are accessible within the Galaxy.  These
signatures allow us to probe back to early epochs.  We believe that the
near-field universe has the same level of importance as the far-field
universe for a comprehensive understanding of galaxy formation and
evolution.

We have argued that understanding galaxy formation is primarily about
understanding baryon dissipation within the CDM hierarchy; to a large
extent, this means understanding the formation of disks. The question
we seek to address is whether this can ever be unravelled in the near
or far field.  Dynamical information was certainly lost at several
stages of this process, but we should look for preserved signatures of
the different phases of galaxy formation.

Far-field cosmology can show how the light-weighted, integrated
properties of disks change with cosmic time. While light-weighted
properties provide some constraint on simulations of the future, they
obscure some of the key processes during dissipation.  The great
advantage of near field studies is the ability to derive ages and
detailed abundances for individual stars within galaxies of the Local
Group.

We have addressed the issue of information content within the
Gaiasphere.  The detailed information that is possible on ages,
kinematics, and chemical properties for a billion stars $-$ which we
see as the limit of observational knowledge over the next two decades
$-$ may reveal vast complexity throughout the disk. It may not be
possible to perceive the sequence of events directly. However, we are
optimistic that future dissipational models may provide unique
connections with the observed complexity.

It is clear that detailed high resolution abundance studies of large
samples of galactic stars will be crucial for the future of fossil
astronomy. Christlieb\etal\ (2000) find that strong r-process enhanced
stars can be identified with $R = 20,000$ and SNR $=$ 30 pix$^{-1}$
from the Eu lines. Both UVES and HDS can reach this sensitivity for a
$B = 15$ star in just 20~min. But the detailed abundance work requires
a substantial increase in the resolving power.  Cayrel\etal\ (2001) and
Hill\etal\ (2002) demonstrate the exquisite quality and capability of
high resolution spectroscopy for CS~31082-001 where they achieve a SNR
$\simeq$ 300 in just four hours with UVES at $R \simeq 60,000$.  See
Fig.~\ref{fig12} for another excellent example.  But these are bright
stars with some of the most extreme overabundances of r-process
elements observed to date.

Gaia will provide accurate distances, ages and space motions for a vast
number of stars, separate with great precision the various Galactic
components, and identify most of the substructure in the outer bulge
and halo.  High resolution spectrographs like UVES on the VLT, HDS on
Subaru, and HIRES on Keck are starting to reveal the rich seam of
information in stellar abundances. 

We must stress that in order to access a representative sample of the
Gaiasphere, this will require a new generation of ground-based
instruments, in particular, a multi-object echelle spectrograph with
good blue response on a large aperture telescope.  We close with a
brief discussion of what is required.

As an example, the FGK sub-giants and giants are a characteristic
population which could be studied over the full extent of the
Gaiasphere, as discussed in the previous section.  Typical stars will
have magnitudes around 17$-$18 which is at the limit of the
state-of-the-art spectrometer UVES at $R \simeq 60,000$.

We now consider what it would take to achieve high resolution
spectroscopy for a representative sample of stars within the
Gaiasphere.  Our baseline instrument UVES achieves cross-dispersed
echelle spectroscopy in two wavelength ranges (300-500nm, 420-1100nm).
For a {\it limiting} resolution of $R \simeq 60,000$ for a single night
exposure, the sensitivity limit is $U \approx 18.0$ and $V \approx
19.5$ in the blue and red arms.  UVES now allows multi-object echelle
spectroscopy (red arm) from fibre inputs provided by the Fibre Large
Array Multi-Element Spectrograph (FLAMES).  This will enable the
simultaneous observation of eight objects over a 25$^\prime$ field of
view.

Existing multi-object spectrographs are mostly used redward of 450nm
because of the fundamental limits of conventional optical fibres.
Normal fibres transmit light through total internal reflection but blue
light is Rayleigh scattered below 450nm. Recently, photonic crystal or
microstructured fibres threaded with air channels (Cregan\etal\ 1999)
have been shown to be highly transmissive down to the atmospheric
cut-off.  This is a technical breakthrough for blue multi-object
spectroscopy.

We believe there is a real need for a high-resolution spectrograph
which can reach hundreds or even thousands of stars in a square degree
or more. The Gemini Wide Field proposal currently under discussion
provides an opportunity for this kind of instrument (S. Barden,
personal communication).  Such an instrument will be expensive and
technically challenging, but we believe this must be tackled if we are
to ever unravel the formation of the Galaxy.

\normalsize
\bigskip
The philosophy behind this review has emerged from discussions dating
back to the spring of 1988 when KCF and JBH were visiting the Institute
of Advanced Study at Princeton. At that time, there was a quorum of
galaxy dynamicists at the IAS whose work continues to inspire and
excite us.  Our thanks go to John Bahcall for this opportunity. We
thank Michael Perryman and the Gaia team for the inspiration of the
Gaia science mission.  We have greatly benefitted from excellent
reviews by E. Friel, J. Sellwood, and G. Wallerstein and
collaborators.  Most recent, we acknowledge the inspiration of
colleagues at the 2001 Dunk Island conference, in particular, Tim de
Zeeuw, Mike Fall, Ivan King, John Kormendy, John Norris, Jerry
Sellwood, Pieter van der Kruit and Ewine van Dishoeck.  We have
benefited from discussions with Vladimir Avila-Reese, Rainer Beck, Bob
Kurucz, Ruth Peterson, Tomek Plewa and Jason Prochaska. We are indebted
to Allan Sandage for many constructive comments.  Finally, we thank the
Editor for suggesting the main title `New Galaxy' for this review.

\newpage
\begin{figure}
\caption{
Look-back time as a function of redshift and the size of the Universe
(Lineweaver 1999) for five different world models. The approximate ages
of the Galactic halo and disk are indicated by hatched regions. 
}
\label{fig1}
\end{figure}
\begin{figure}
\caption{
The age-metallicity relation of the Galaxy for the different components
(see text): TDO $-$ thin disk open clusters;
TDG $-$ thick disk globulars; B $-$ bulge;
YHG $-$ young halo globulars; OHG $-$ old halo globulars.  The blue
corresponds to thin disk field stars, the green to thick disk field
stars, and the black shows the distribution of halo field stars extending
down to [Fe/H] = -5.  
} 
\label{fig2} 
\end{figure}

\begin{figure}
\caption{
(a) Sketch of Milky Way showing the stellar disk (light blue), thick
disk (dark blue), stellar bulge (yellow), stellar halo (mustard yellow),
dark halo (black) and globular cluster system (filled circles). The radius
of the stellar disk is roughly 15 kpc.  The baryon and dark halos
extend to a radius of at least 100 kpc.
(b)  Infrared image of the Milky Way taken by the DIRBE instrument on board
the Cosmic Background Explorer (COBE) Satellite. [We acknowledge the NASA 
Goddard Space Flight Center and the COBE Science Working Group for this
image.]
(c) M104, a normal disk galaxy with a large stellar bulge (from AAO).
(d) Hubble Heritage image of the compact group Hickson 87; one galaxy
has a peanut-shaped stellar bulge due to dynamical interaction with other
group members.
(e) Image of the SO galaxy NGC 4762 (Digital Sky Survey) shows its thin disk 
and stellar bulge.
(f) A deeper image of NGC 4762 (DSS) shows its more extended thick
disk. The base of the arrows in (e) and (f) shows the height above the plane
at which the thick disk becomes brighter than the thin disk (Tsikoudi 1980).
(g) Image of the pure disk galaxy IC 5249 (DSS) shows its thin disk and no
stellar bulge.
(h) A deeper image of IC 5249 (DSS) shows no visible thick disk,
although a very faint thick disk has been detected in deep surface
photometry.
}
\label{fig3}
\end{figure}
\begin{figure}
\caption{
The density distribution of Sa to Sm galaxies over effective
radius $r_e$ and effective surface brightness $\mu_e$. The
top panel shows the raw (unweighted) distribution and the bottom
panel shows the luminosity weighted distribution (de~Jong
\& Lacey 2000).
}
\label{fig4}
\end{figure}
\begin{figure}
\caption{
The relation between the three components of the velocity dispersion and
the stellar age, as derived by Quillen \& Garnett (2001) for stars from the 
sample of Edvardsson \etal\ (1993). Stars with ages between $2$ and $10$ Ga
belong to the old thin disk: their velocity dispersion is independent of age.
The younger stars show a smaller velocity dispersion. The velocity dispersion
doubles abruptly at an age of about 10 Ga: these older stars belong to the 
thick disk. 
}
\label{fig5}
\end{figure}
\begin{figure}
\caption{
Sauron integral field observations of NGC 4365 (top panels) and NGC
4150 (bottom panels).  Left panels: reconstructed surface brightness
maps.  Right panels: H$\beta$ versus [MgFe5270] diagram.  The points
were derived from the Sauron datacubes by averaging along the
corresponding color-coded isophotes (Bacon\etal\ 2001; Davies\etal\
2001). The [Fe/H] vs. age grid is derived from Vazdekis (1999).  
[We acknowledge Harald Kuntschner and the Sauron team for these
images.]
}
\label{fig6} 
\end{figure}
\begin{figure}
\caption{
Examples of normal spirals with faint stellar streamers in the outer 
halo (see text):
(a) M104 where the streamer is on a much larger scale
than shown in Fig.~3(c) (Malin \& Hadley 1997);
(b) M83 (Malin \& Hadley 1997);
(c) NGC 5907 (Shang\etal\ 1998);
(d) M31 (Ibata\etal\ 2001$a$).
}
\label{fig7}
\end{figure}
\begin{figure}
\caption{
(a) The distribution of open clusters younger than Hyades with height from 
the plane as a function of Galactocentric distance $R_{\rm gc}$ (Friel 1995). 
The Sun is at 8.5~kpc. (b) The distribution of clusters with ages equal 
to or greater than the Hyades.  (c) The open clusters exhibit a well defined 
abundance gradient. (d) There is no discernible age-metallicity relation (AMR) 
when the cluster abundances are corrected for the radial abundance gradient.
}
\label{fig8}
\end{figure}
\begin{figure}
\caption{
A satellite in orbit about the Milky Way as it
would appear after 8 Ga. While stars from the disrupted satellite
appear to be dispersed over a very wide region of sky, it will be
possible to deduce the parameters of the original event using special
techniques (see text).  [We acknowledge A. Helmi and S. White for
this image.]
}
\label{fig9}
\end{figure}
\begin{figure}
\caption{
(a) Initial distribution of particles for 33 systems
falling into the Galactic halo is integral of motion space.
(b) The final distribution of particles in (a) after 12~Ga; the
data points have been convolved with the errors expected for Gaia.  
[We acknowledge A. Helmi for these images.]
A simulation of the baryon halo built up through accretion of
100 satellite galaxies. (c) The different colors show the disrupted
remnants of individual satellites. (d) This is the same simulation
shown in a different coordinate frame, i.e. the orbit radius
(horizontal) plotted against the observed radial velocity (vertical)
of the star. [We acknowledge P. Harding and H. Morrison for these
images.]
}
\label{fig10}
\end{figure}
\begin{figure}
\caption{
Mean relative abundance ratios of light s-process elements (top
panel), heavy s-process elements (middle panel), and and r-process
elements (bottom panel) as functions of [Fe/H]. In each panel,
the dotted horizontal lines represent the solar abundance ratios of
these elements. The references for the data points are given in
Wallerstein\etal\ (1997). [We acknowledge C. Sneden for this figure.]
}
\label{fig11}
\end{figure}
\begin{figure}
\caption{
CS 22892-052 n-capture abundances (points) taken from Sneden\etal\
(2000) and {\it scaled} solar system abundances  (solid and dashed lines)
taken from Burris\etal\ (2000). Many of the heavy elements conform to
the solar system r-process abundance pattern, although some elements show
the hallmark of the s-process. This figure was originally presented
in Sneden\etal\ (2001$a$).
}
\label{fig12}
\end{figure}

\end{document}